\documentclass[iop]{emulateapj}

\usepackage{natbib}

\newcommand{\kms}{km\,s$^{-1}$}
\newcommand{\cii}{[\ion{C}{2}]}
\newcommand{\mgii}{\ion{Mg}{2}}
\newcommand{\lfir}{$L_{\mathrm{FIR}}$}
\newcommand{\ltir}{$L_{\mathrm{TIR}}$}
\newcommand{\lcii}{$L_\mathrm{[CII]}$}
\newcommand{\lbol}{$L_\mathrm{bol}$}
\newcommand{\lsun}{$L_\sun$}
\newcommand{\msun}{$M_\sun$}
\newcommand{\mdyn}{$M_\mathrm{dyn}$}
\newcommand{\mbh}{$M_\mathrm{BH}$}
\newcommand{\msunyr}{$M_\sun$\,yr$^{-1}$}

\newcommand{\jykms}{Jy\,km\,s$^{-1}$}
\newcommand{\vcirc}{$v_\mathrm{circ}$}

\slugcomment{Accepted for publication in ApJ}

\shorttitle{\cii\ and dust emission in $z>6.6$ quasar hosts}
\shortauthors{Venemans et al.}

\begin{document}

\title{Bright \cii\ and dust emission in three $z>6.6$ quasar host
  galaxies observed by ALMA}

\author{Bram P.\ Venemans\altaffilmark{1},
Fabian Walter\altaffilmark{1},
Laura Zschaechner\altaffilmark{1},
Roberto Decarli\altaffilmark{1},
Gisella De Rosa\altaffilmark{2,3,4},
Joseph R.\ Findlay\altaffilmark{5},
Richard G. McMahon\altaffilmark{6,7},
Will J.\ Sutherland\altaffilmark{8}
}
\altaffiltext{1}{Max-Planck Institute for Astronomy, K{\"o}nigstuhl 17, 69117
  Heidelberg, Germany}
\email{venemans@mpia.de}
\altaffiltext{2}{Department of Astronomy, The Ohio State University,
  140 West 18th Avenue, Columbus, OH 43210, USA}
\altaffiltext{3}{Center for Cosmology and AstroParticle Physics, The
  Ohio State University, 191 West Woodruff Ave, Columbus, OH 43210,
  USA}
\altaffiltext{4}{Space Telescope Science Institute, 3700 San Martin
  Drive, Baltimore, MD 21218, USA}
\altaffiltext{5}{Department of Physics and Astronomy, University of
  Wyoming, Laramie, WY 82071, USA}
\altaffiltext{6}{Institute of Astronomy, University of Cambridge,
  Madingley Road, Cambridge CB3 0HA, UK}
\altaffiltext{7}{Kavli Institute for Cosmology, University of
  Cambridge, Madingley Road, Cambridge CB3 0HA, UK}
\altaffiltext{8}{School of Physics and Astronomy, Queen Mary
  University of London, Mile End Road, London, E1 4NS, UK}

\begin{abstract}
We present ALMA detections of the \cii\ 158\,$\mu$m emission line and
the underlying far-infrared continuum of three quasars at $6.6<z<6.9$
selected from the VIKING survey. The \cii\ line fluxes range between
1.6--3.4\,Jy\,\kms (\cii\ luminosities
$\sim$$(1.9-3.9)\times10^9$\,\lsun). We measure continuum flux
densities of 0.56--3.29\,mJy around 158\,$\mu$m (rest-frame), with
implied far-infrared luminosities between
$(0.6-7.5)\times10^{12}$\,\lsun\ and dust masses
$M_d=(0.7-24)\times10^8$\,\msun. In one quasar we derive a dust
temperature of $30^{+12}_{-9}$\,K from the continuum slope, below the
canonical value of 47\,K. Assuming that the \cii\ and continuum
emission are powered by star formation, we find star-formation rates
from 100--1600\,\msunyr\ based on local scaling relations. The
\lcii/\lfir\ ratios in the quasar hosts span a wide range from
$(0.3-4.6)\times10^{-3}$, including one quasar with a ratio that is
consistent with local star-forming galaxies. We find that the strength
of the \lcii\ and 158\,$\mu$m continuum emission in $z\gtrsim6$ quasar
hosts correlate with the quasar's bolometric luminosity. In one
quasar, the \cii\ line is significantly redshifted by
$\sim$1700\,\kms\ with respect to the \mgii\ broad emission line.
Comparing to values in the literature, we find that, on average, the
\mgii\ is blueshifted by 480\,\kms\ (with a standard deviation of
630\,\kms) with respect to the host galaxy redshift, i.e.\ one of our
quasars is an extreme outlier. Through modeling we can rule out a flat
rotation curve for our brightest \cii\ emitter. Finally, we find that
the ratio of black hole mass to host galaxy (dynamical) mass is higher
by a factor 3--4 (with significant scatter) than local relations.
\end{abstract}


\section{INTRODUCTION}

One of the outstanding questions in astronomy is when the first
galaxies formed, and what their physical properties were. In recent
years, enormous progress has been made in finding galaxy candidates up
to $z\gtrsim10$, $\sim$450 million years after the Big Bang
\citep[e.g.,][]{rob10,mad14}. However, the extremely faint magnitudes
(observed infrared magnitudes $J_{\mathrm{AB}}\gtrsim26$) and modest
star formation rates (SFRs $\lesssim10$ \msunyr) of these early
galaxies make it nearly impossible to study the properties of the
stars, gas and dust in much detail, even with current state-of-the-art
facilities, including ALMA. An effective way to learn more about the
constituents of galaxies at the highest redshifts is to study the
brightest (and most massive) members of this population ($L \gg
L^*$). Such bright galaxies are very rare, and not found in the deep,
pencil-beam searches typically used for high-$z$ galaxy searches,
e.g.\ with the {\em Hubble Space Telescope}.

\begin{deluxetable*}{lccc}
\tablecaption{Description of the ALMA observations \label{tab:obsdesc}}
\tablewidth{0pt} 
\tablehead{ \colhead{} & \colhead{J2348--3054} & \colhead{J0109--3047} &
  \colhead{J0305--3150}}
\startdata
R.A. (J2000) & 23$^h$48$^m$33$^s\!\!$.35 & 01$^h$09$^m$53$^s\!\!$.13 &
03$^h$05$^m$16$^s\!\!$.91 \\
Decl. (J2000) & --30$^\circ$54\arcmin10\farcs28 &
--30$^\circ$47\arcmin26\farcs32 &  --31$^\circ$50\arcmin55\farcs94 \\
$z_\mathrm{MgII}$\tablenotemark{a} & $6.889_{-0.006}^{+0.007}$ &
$6.747_{-0.005}^{+0.007}$ & $6.605_{-0.001}^{+0.002}$ \\
$\nu_\mathrm{obs}$ (GHz) & 240.575 & 245.231 & 249.841 \\
$t_\mathrm{exp,on-source}$ (min) & 17 & 16 & 16 \\
\# of antennas & 18--30 & 23 & 23 \\
RMS noise (per 100\,MHz) & 0.44\,mJy & 0.48\,mJy & 0.29\,mJy \\
beam size & 0\farcs74\,$\times$\,0\farcs54 & 0\farcs70\,$\times$\,0\farcs45 
& 0\farcs62\,$\times$\,0\farcs44
\enddata
\tablenotetext{a}{Taken from \citet{der14}}
\end{deluxetable*}

In the local universe, it is argued that a tight correlation between
the mass of a galaxy and the black hole that it harbors exist
(e.g.\ see recent review by \citealt{kor13}). Such a correlation seems
to be in place also in the high redshift universe, at least to first
order, as the host galaxies of bright quasars at $z>2$ (powered by
supermassive, $>$10$^9$\,\msun\ black holes) are among the brightest
and most massive galaxies found at these redshifts
\citep[e.g.,][]{sey07,deb10,morr12}. Therefore, an effective method to
pinpoint the most massive and luminous galaxies in the early universe
is believed to locate bright quasars at the highest redshifts. The
Sloan Digital Sky Survey (SDSS) discovered $\sim$30 bright ($M_{1450}
< -26$) quasars around $z\sim6$ which are shown to host supermassive,
$>$10$^9$\,\msun\ black holes
\citep[e.g.,][]{fan03,fan06b,jia07,kur07,der11}. Observations of the
host galaxies of these quasars in the radio and (sub)mm demonstrated
that large reservoirs of dust and metal enriched atomic and molecular
gas can exist in massive galaxies up to $z\sim6.4$, less than 1\,Gyr
after the Big Bang
\citep[e.g.,][]{ber03a,ber03b,wal03,mai05,wan11b,wan13}. These
observations already provide constraints on models of massive galaxy
and dust formation at high redshift, requiring large initial gas
masses and efficient supernova dust production
\citep[e.g.,][]{mai04,mic10,gal11,kuo12,val14}.

To further constrain the build-up of massive galaxies, the growth of
supermassive black holes and the formation of dust in the early
universe, it is important to locate and study bright quasars at the
highest redshifts possible. Over the last four years, we have
discovered seven quasars with redshifts above $z>6.5$ (the SDSS limit)
using wide-field near-infrared surveys with redshifts up to $z=7.1$
\citep{mor11,ven13,ven15a}. These 7 quasars are currently the only
known quasars at $z>6.5$. These new $z>6.5$ quasars are as bright as
quasars at $z\sim6$ and are powered by black holes with masses in
excess of $\gtrsim10^9$\,\msun\ \citep{mor11,ven13,ven15a,der14},
constraining models of black hole formation \citep[e.g.,][]{der14}.

In this paper, we report the detection of bright \cii\ and dust
continuum emission in three quasars at $z>6.6$. These are VIKING
J234833.34--305410.0 (hereafter J2348--3054), VIKING
J010953.13--304726.3 (hereafter J0109--3040), and VIKING
J030516.92--315056.0 (hereafter J0305--3150), discovered in
\citet{ven13}. The paper is organized as follows. In
Section~\ref{sec:observations} we describe the ALMA Cycle 1
observations. In Section~\ref{sec:results} we present our results: in
Section~\ref{sec:luminosities} we provide the detailed luminosities
for each source, followed by a description of additional sources in
the quasars fields in Section~\ref{sec:companions}. In
Section~\ref{sec:discussion} we discuss our findings: firstly, in
Section~\ref{sec:ciifir} we compare the \cii/\lfir\ ratios of $z>6.5$
quasar hosts with lower redshift counterparts, followed by a
discussion of possible correlations between optical/near-infrared and
far-infrared properties of high redshift quasars in
Section~\ref{sec:correlations}. In Section~\ref{sec:ciimgii} we look
into the difference in redshift given by the rest-frame UV \mgii\ and
far-infrared \cii\ lines. In Section~\ref{sec:j0305} we investigate
the properties of the source detected with the highest significance,
J0305--3150, followed by a discussion on the effects of the cosmic
microwave background (CMB) on the observations in
Section~\ref{sec:cmb}. In Section~\ref{sec:mdyn} we estimate dynamical
masses of the quasar host galaxies using the detected \cii\ lines. We
conclude with a summary in Section~\ref{sec:summary}.

Throughout this paper, we adopt the following cosmological parameters:
$H_0=70$ km\,s$^{-1}$\,Mpc$^{-1}$, $\Omega_M=0.28$, and
$\Omega_\lambda=0.72$ \citep{kom11}. Star formation rates (SFRs) are
calculated assuming a \citet{kro03} initial mass function (IMF).

\section{ALMA OBSERVATIONS}
\label{sec:observations}

Observations of the three $6.6<z<6.9$ quasars were carried out between 2013
July 5 and 2013 November 14. The setup of the observations for each of
the sources was to have two overlapping sidebands covering the
\cii\ line (tuned using the \mgii\ redshift). The expected frequency of
the \cii\ line was 240.9, 245.3, and 250.1\,GHz for J2348--3054,
J0109--3047, and J0305--3150, respectively. The overlap between the
two bandpasses was 20\%, leaving a total frequency coverage of
3.375\,GHz around the expected frequency of the \cii\ line. At
$z\sim6.7$, this frequency coverage corresponds to
$\sim$4100\,\kms. The other two bandpasses were placed at
approximately 15\,GHz (observed) below the \cii\ frequency to measure
the far-infrared (FIR) continuum.

The observations were carried out in a compact configuration
(baselines below 1\,km and mostly below 300\,m). The number of
antennas used varied between 18 in July 2013 to 30 in November
2013. Bandpass calibration was performed through observations of
J0334--4008, J0522--3627, and J2258--2758. For the flux and amplitude
calibration, the sources J2357--5311, J0334--401, and Neptune were
observed. The pointing was checked on sources J0120--2701,
J0334--4008, J2357--5311, J0522--3627, and J2258--2758. Finally, the
phase calibrators J0120--2701, J0334--4008, and J2339--3310 were
observed every 7--8\,min. The total on-source integration times on the
quasars were 16--17\,min per source.

Standard reduction steps using the Common Astronomy Software
Applications package (CASA) were followed. Some flagging due to
atmospheric lines was required, although such flagging was kept to a
minimum in order to keep as much of the bandwidth as possible. Minimal
additional flagging was required. High-frequency striping was present
in the data, which was eliminated by removing the long
baselines. Self-calibration was attempted with and without the long
baselines, but showed no additional improvement and thus was not used
for the final cubes. The cubes were cleaned using a weighting factor
of robust=0.5, which optimized the noise per frequency bin and the
resolution of the resulting map.

A summary of the observation is provided in Table \ref{tab:obsdesc}.

\section{RESULTS}
\label{sec:results}

\begin{figure}
\includegraphics[width=\columnwidth]{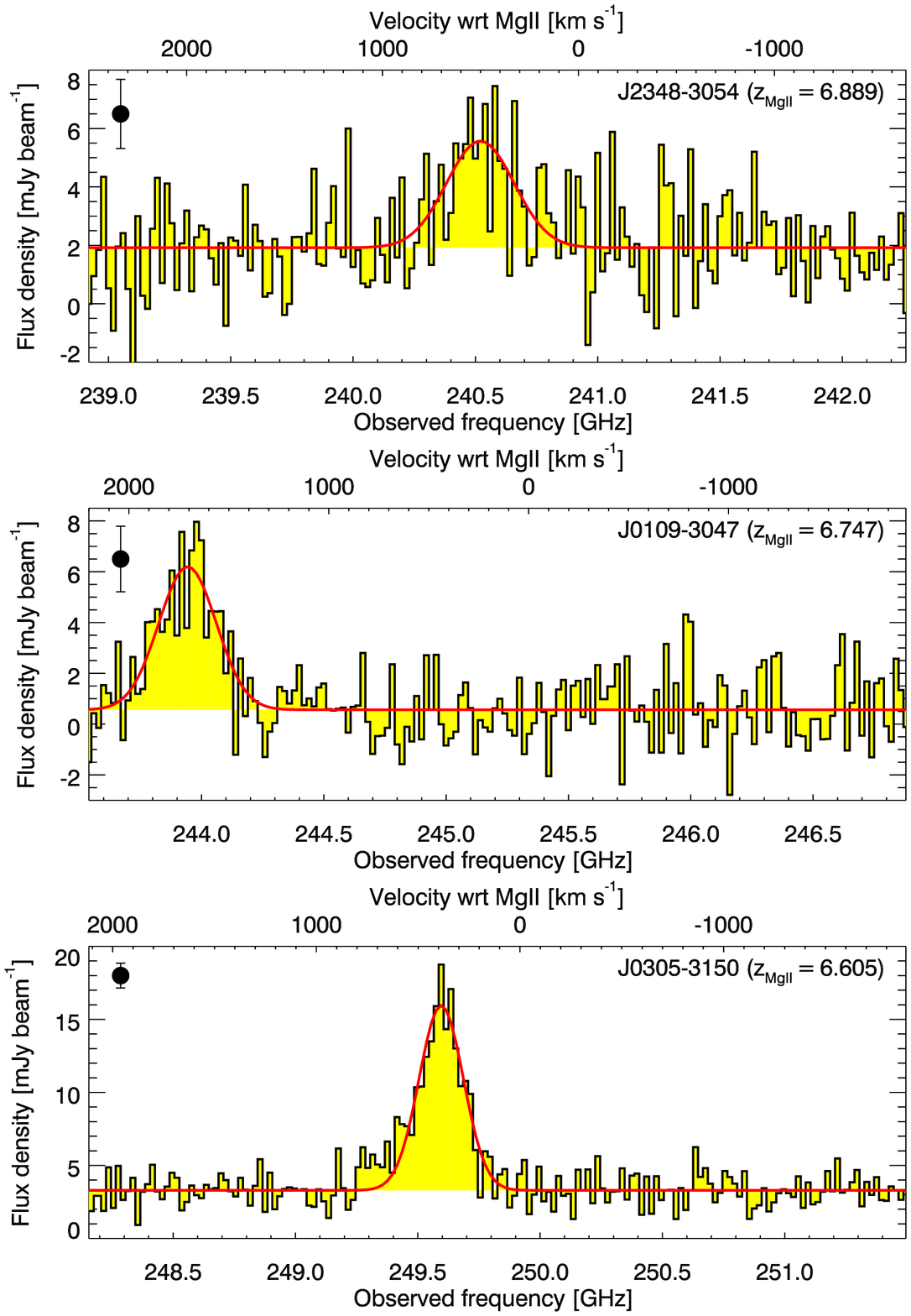}
\caption{\cii\ spectra of the three $z>6.6$ quasars observed with
  ALMA. The spectra were extracted from the data cubes smoothed with a
  1\arcsec\ Gaussian at the location of the brightest pixel in the
  emission line map (Fig.~\ref{fig:maps}), which in all cases
  coincides with the optical/near-infrared position of the
  quasars. Only the two bandpasses encompassing the emission line are
  shown. The bottom axis shows the observed frequency in GHz and on
  the top we plot the velocity with respect to the redshift of the
  \mgii\ line, which is also given in the top right corner of each
  spectrum. The solid line represents a Gaussian+continuum fit to the
  data. The typical uncertainty per bin is plotted in the upper left
  corner of each spectrum. \label{fig:spectra}}
\end{figure}

\begin{figure}
\includegraphics[width=\columnwidth]{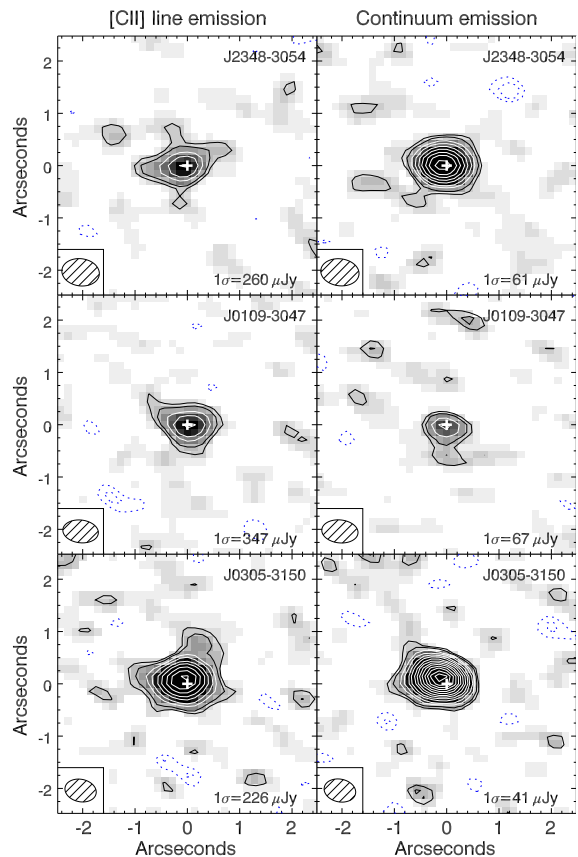}
\caption{Maps of the line emission (left) and continuum (right) of the
  VIKING quasars. For the line maps the line emission was averaged
  over the FWHM, measured from the spectrum of the central pixel
  (Fig.~\ref{fig:spectra} and Table~\ref{tab:firprop}), and the
  continuum emission was subtracted. The beam is shown in the bottom
  left of each map. The emission was averaged over 450\,\kms,
  330\,\kms, and 225\,\kms\ in the case of J2348--3054, J0109--3047,
  and J0305--3150, respectively. The $1\sigma$ rms noise of each map
  is printed at the bottom right. The small white cross indicates the
  optical/near-infrared position of the quasar. The blue, dashed
  contours are $-3\sigma$ and $-2\sigma$. The black, solid contours
  are $+2\sigma$ and $+3\sigma$, the white solid contours are [5, 7,
    10, 13, 17, 21, 26, 31, 37, 43, 50, 57]\,$\times\sigma$.
  \label{fig:maps}}
\end{figure}

\begin{deluxetable*}{lccc}
  \tablecaption{Far-infrared Properties of The $z>6.6$ Quasars
    \label{tab:firprop}}
\tablewidth{0pt} 
\tablehead{ \colhead{} & \colhead{J2348--3054~} & \colhead{J0109--3047~} &
  \colhead{J0305--3150~}}
\startdata
\cii\ redshift & $6.9018\pm0.0007$ & $6.7909\pm0.0004$ & $6.6145\pm0.0001$ \\
\cii-\mgii\ velocity shift [km\,s$^{-1}$] & $486\pm267$ & $1690\pm232$~ &
$374\pm79$~ \\
\cii\ line width (FWHM) [km\,s$^{-1}$] & $405\pm69$~ & $340\pm36$~ &
$255\pm12$~ \\
\cii\ line flux [Jy\,km\,s$^{-1}$] & $1.57\pm0.26$ & $2.04\pm0.20$ & 
$3.44\pm0.15$ \\
Continuum flux density\tablenotemark{a} [mJy] & $1.92\pm0.14$ & 
$0.56\pm0.11$ & $3.29\pm0.10$ \\
\cii\ equivalent width (EW$_\mathrm{[CII]}$) [$\mu$m] & $0.43\pm0.08$
& $1.90\pm0.42$ & $0.55\pm0.03$ \\
\cii\ luminosity [$10^9$\,$L_\sun$] & $1.9\pm0.3$ & $2.4\pm0.2$ & $3.9\pm0.2$ \\
FIR luminosity [$10^{12}$\,$L_\sun$] & $2.4-4.9$ & $0.6-1.5$ &
~$4.0-7.5$\tablenotemark{b} \\
TIR luminosity [$10^{12}$\,$L_\sun$] & $3.8-6.9$ & $0.9-2.2$ &
~~$6.3-10.6$\tablenotemark{b} \\
\lcii/\lfir & $(0.33-0.94)\times10^{-3}$ & $(1.4-4.6)\times10^{-3}$ &
$(0.50-1.03)\times10^{-3}$\tablenotemark{b} \\
SFR$_\mathrm{TIR}$ [\msunyr] & ~$555-1020$ & $140-325$ &
~$940-1580$\tablenotemark{b} \\
SFR$_\mathrm{[CII]}$ [\msunyr] & $100-680$ & $140-895$ & ~$250-1585$ \\
$M_\mathrm{dust}$ [$10^8$\,\msun] & $2.7-15$ & $0.7-4.9$ & $4.5-24$ \\
Deconvolved size \cii & $<$0\farcs74\,$\times$\,0\farcs54\tablenotemark{c} & 
              (0\farcs43$\pm$0\farcs10)\,$\times$\,(0\farcs39$\pm$0\farcs15) &
              (0\farcs60$\pm$0\farcs03)\,$\times$\,(0\farcs42$\pm$0\farcs04) \\
Deconvolved size \cii [kpc] & $<$4.0\,$\times$\,2.9\tablenotemark{c} & 
              (2.3$\pm$0.5)\,$\times$\,(2.1$\pm$0.8) &
              (3.3$\pm$0.2)\,$\times$\,(2.3$\pm$0.2) \\
Deconvolved size continuum & $<$0\farcs76\,$\times$\,0\farcs53\tablenotemark{c}
            & $<$0\farcs71\,$\times$\,0\farcs45\tablenotemark{c} & 
              (0\farcs40$\pm$0\farcs02)\,$\times$\,(0\farcs29$\pm$0\farcs02) \\
Deconvolved size continuum [kpc] & $<$4.1\,$\times$\,2.9\tablenotemark{c} & 
              $<$3.9\,$\times$\,2.5\tablenotemark{c} &
              (2.2$\pm$0.1)\,$\times$\,(1.6$\pm$0.1)
\enddata
\tablenotetext{a}{Continuum flux density at a rest-frame wavelength of 
158\,$\mu$m.}
\tablenotetext{b}{In Section~\ref{sec:cmb} we obtain $T_d=30_{-9}^{+12}$\,K
  by fitting the continuum slope of J0305--3150 while taking the effects of the cosmic microwave background into account. With a dust
  temperature of $T_d=30$\,K, we derive \lfir\,$=2.6\times10^{12}$\,\lsun,
  \ltir\,$=3.7\times10^{12}$\,\lsun, \lcii/\lfir\,$=1.5\times10^{-3}$,
  and SFR$_\mathrm{TIR}=545$\,\msunyr.}
\tablenotetext{c}{Unresolved}
\end{deluxetable*}

\subsection{[CII] and FIR emission}
\label{sec:luminosities}

All three quasars are detected in the ALMA data in both the continuum
and line emission. In Fig.\ \ref{fig:spectra} we show the spectrum of
the brightest pixel in the spectral regions encompassing the
\cii\ line after smoothing the data cubes with a 1\arcsec\ Gaussian
using the CASA task `imsmooth'. We fitted a Gaussian + constant to the
spectra to model the \cii\ emission and the continuum. The Gaussian
fit provided the redshift, width, and strength of the emission line,
which are listed in Table \ref{tab:firprop}. We averaged the continuum
subtracted data cubes over the FWHM around the line centre to produce
a map of the line emission (Fig.\ \ref{fig:maps}). The channels in
bandpasses 0 and 1 that did not contain line emission, and bandpasses
2 and 3 were averaged to create maps of the continuum emission, which
are also shown in Fig.~\ref{fig:maps}. From the maps we measured the
sizes of the line and continuum emission using CASA task `imfit'.

\begin{figure}
\includegraphics[width=\columnwidth]{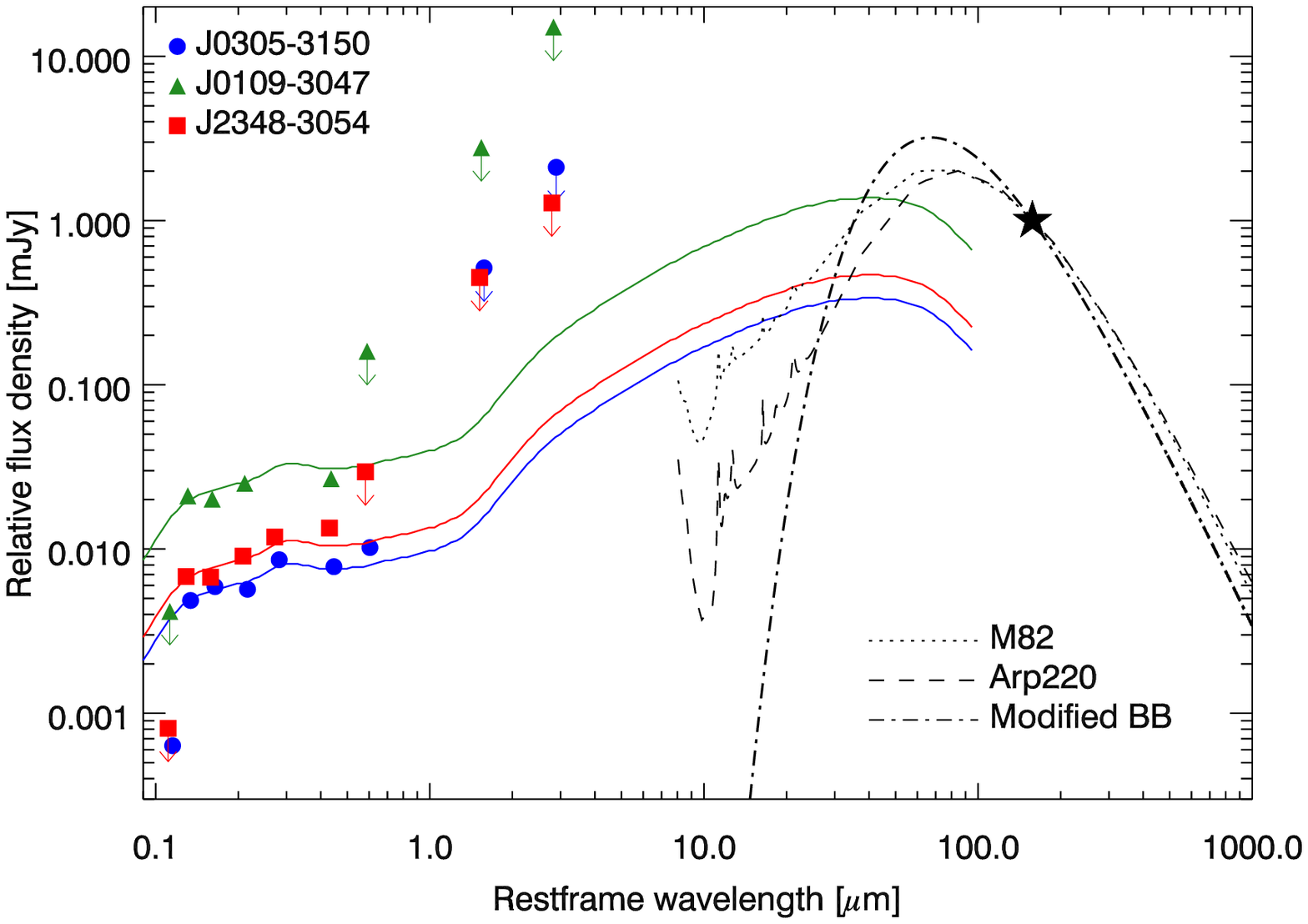}
\caption{Spectral energy distributions (SEDs) of the three VIKING
  quasars, normalized to 1\,mJy at 158\,$\mu$m in the rest-frame. The
  rest-frame UV to near-infrared data points are taken from
  \citet{ven13} and from the {\it Wide-field Infrared Survey Explorer}
  \citep[{\it WISE},][]{wri10}. We fitted the quasar template of
  \citet{ric06} to the short wavelength
  ($\lambda_\mathrm{obs}<30$\,$\mu$m) data points of each quasar. The
  three models used in this paper to model the far-infrared emission
  are shown by the dotted (M82 template), dashed (Arp 220 template),
  and dot-dashed (modified black body) lines. \label{fig:sed}}
\end{figure}

To derive far-infrared properties of the quasar hosts, we applied the
same assumptions as \citet{ven12}. In short, the far-infrared
luminosity (\lfir) is defined as the luminosity integrated from
42.5\,$\mu$m to 122.5\,$\mu$m in the rest-frame
\citep[e.g.,][]{hel88}. The total infrared luminosity (\ltir) was
computed by integrating the continuum from 8\,$\mu$m to 1000\,$\mu$m
in the rest-frame. For the shape of the infrared continuum, we assumed
three different models. The first is a modified black body: $f_\nu
\propto B_\nu(T_d)(1-e^{-\tau_d})$ with $B_\nu$ the Planck function with
a dust temperature $T_d$ and $\tau_d$ the dust optical depth
\citep[e.g.,][]{bee06}. Following the literature, our modified black
body model has a dust temperature of $T_d=47$\,K and a dust emissivity
power-law spectral index of $\beta=1.6$ \citep[see
  e.g.,][]{bee06,lei13}. We further assume that the dust optical depth
is low at far-infrared wavelengths, $\tau \ll 1$, at $\lambda >
40$\,$\mu$m. The other two models are templates of the local
star-forming galaxies Arp 220 and M82 \citep{sil98}. Note that while
the dust temperature of Arp 220 is found to be higher than the
temperature assumed for our modified black body model,
$T_d(\mathrm{Arp} 220)=66$\,K, the dust opacity is also higher with
$\tau_d\approx2$ at 158\,$\mu$m \citep[e.g.,][]{ran11}. We show the
three different templates, combined with the rest-frame UV and optical
photometry of the quasars, in Fig.~\ref{fig:sed}. We caution that the
range of values of \lfir\ and \ltir\ for the VIKING quasar hosts
presented here strongly depends on our choice of models, see also
Section~\ref{sec:cmb}. Additional far-infrared photometry is required
to better constrain the shape of the infrared continuum and thus
\lfir\ and \ltir.

If we assume that the continuum flux density measured around
158\,$\mu$m arises from star formation (which seems to be a valid
assumption for FIR detected quasars at $z>5$, see e.g.,
\citealt{lei14,bar15}), then we can use the local scaling relation
between SFR and \ltir\ from \citet{mur11} to obtain a measurement of
the SFR in the quasar host:
SFR$_\mathrm{TIR}$/\msunyr\,$=3.88\times10^{-44}$\,\ltir/erg\,s$^{-1}$.
Alternatively, we can use the \cii\ emission to calculate the SFR by
applying the relation between \cii\ luminosity and SFR found by
\citet[][]{del14} for high redshift ($0.5<z\lesssim6$) galaxies:
SFR$_\mathrm{[CII]}$/\msunyr\,=\,$3.0\times10^{-9}$\,(\lcii/\lsun)$^{1.18}$,
with an uncertainty of 0.4\,dex. Using instead the relation between
SFR and \lcii\ derived by \citet{del11} and \citet{sar14} the SFRs
would be a factor $\sim$2--2.5 lower. The reason for this difference
is that the latter relations are derived for star-forming galaxies
with SFRs below $\lesssim$100\,\msunyr\ and FIR luminosities
\lfir\,$\lesssim10^{12}$\,\lsun\ and might not be applicable for our
high redshift, \lfir\,$\gtrsim10^{12}$\,\lsun\ quasar hosts (see,
e.g., the discussion in \citealt{del14}). Similarly, if we apply the
relation derived by \citet{her15} for 46 local galaxies with \ltir\,$<
10^{11}$\,\lsun, then the resulting SFR$_\mathrm{[CII]}$ are a factor
$\sim$5--6 lower. They suggest that sources with $10^{11}<$\,\ltir\,$<
10^{12}$\,\lsun\ have a relation that is a factor 1.9 higher, which
would give roughly similar SFRs as \citet{del11} and
\citet{sar14}. Finally, we derived total dust masses both by using the
M82 and Arp220 templates and by assuming a dust temperature of 47\,K
and a dust mass opacity coefficient of $\kappa_\lambda=0.77\,(850\,
\mu\mathrm{m}/\lambda)^\beta$\,cm$^2$\,g$^{-1}$ \citep{dun00}. Since
the dust temperatures in these quasar host is assumed to be
significantly higher than the temperature of the CMB at these redshifts,
$T_\mathrm{CMB}(z=6.7)\approx21$\,K, we ignore the effect of the CMB
on the ALMA observations in
Sections~\ref{sec:j2348res}--\ref{sec:j0305res} \citep[but
  see][]{dac13}. We will however further address the effects of the
CMB in Section~\ref{sec:j0305cont}. The results are also summarized in
Table \ref{tab:firprop}.

\begin{figure*}
\includegraphics[width=\textwidth]{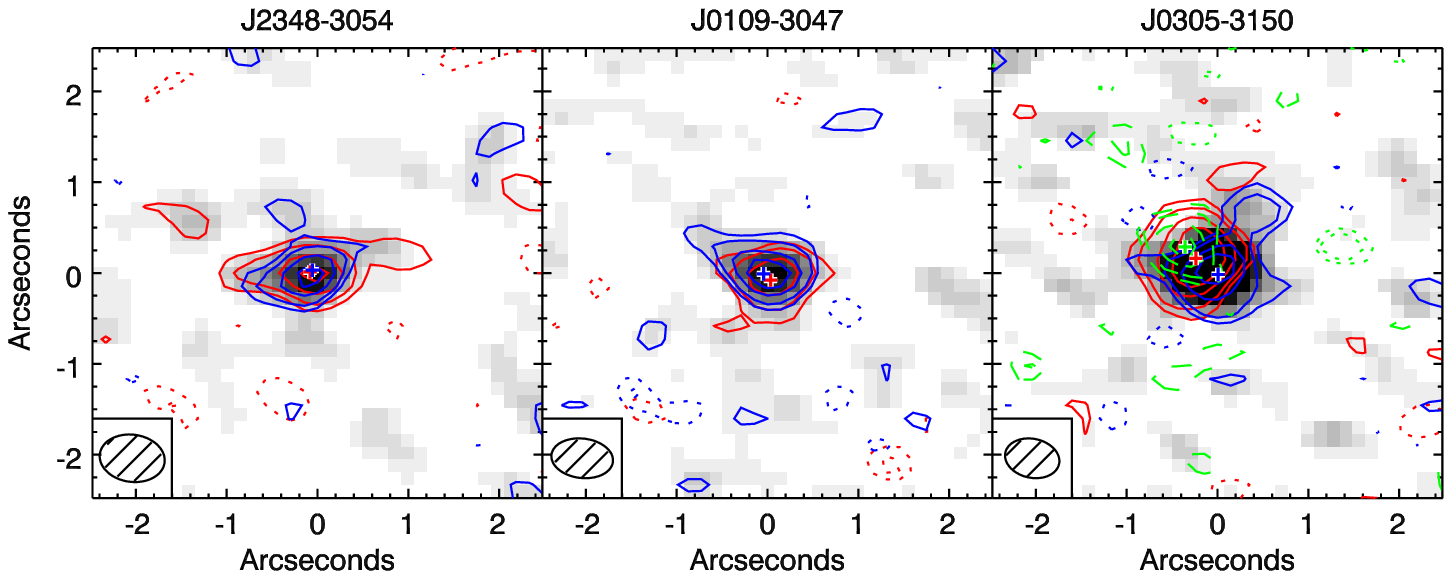}
\caption{Contour plots of the line emission averaged of the FWHM (in
  grey scale) and the blue side and red side of the emission line (in
  blue and red contours). Contour levels are [$-3$, $-2$, +2, +3, +5,
    +7, +10]\,$\times\sigma$. In the case of J2348--3054 and
  J0109--3047 the blue side of the line was centered +81.3\,MHz and
  +69.2\,MHz (+FWHM/4) from the frequency of the line center and
  averaged over 162.6\,MHz and 138.5\,MHz (FWHM/2), respectively. The
  red side was centered $-$81.3\,MHz and $-$69.2\,MHz from the line
  center. For J0305--3150, the maps of the blue and red side of the
  line were constructed by averaging over 141.7\,MHz
  (2/3\,$\times$\,FWHM) centered +141.7\,MHz and $-$141.7\,MHz,
  respectively. For J0305--3150 we also show a map in green,
  long-dashed contours of the emission averaged from 249.25\,GHz to
  249.39\,GHz. At these frequencies there appears to
  be an excess of emission over the Gaussian fit (see
  Fig.~\ref{fig:spectra}). We will discuss this emission in
  Section~\ref{sec:j0305wing}. No offsets between the red and blue
  emission are found for the quasars J2348--3054 and J0109--3047, but
  J0305--3150 shows indications for intrinsic gas motions on the
  scales resolved here (see also
  Section~\ref{sec:dynmodels}). \label{fig:velmaps}}
\end{figure*}

\subsection{J2348--3054}
\label{sec:j2348res}

J2348--3054 is the highest redshift quasar of our sample, with
$z_\mathrm{MgII}=6.889$ \citep{ven13,der14}. The \cii\ emission line
is detected with a peak signal-to-noise S/N\,$\sim10$ at
$z_\mathrm{[CII]}=6.9018\pm0.0007$ (Fig.~\ref{fig:maps}). The emission
line has a peak flux density of $f_p=3.64\pm0.52$\,mJy\,beam$^{-1}$
and a FWHM of $405\pm69$\,\kms. The line emission is unresolved within
the 0\farcs74$\times$0\farcs54 beam (see also
Fig.~\ref{fig:velmaps}). The integrated line flux derived from the
Gaussian fit to the spectrum (Fig.~\ref{fig:spectra}) is
$F_\mathrm{[CII]}=1.57\pm0.26$\,\jykms\ which corresponds to a
luminosity of \lcii\,$=(1.9\pm0.3)\times10^9$\,\lsun, approximately
two times brighter than the $z=7.1$ quasar J1120+0641 \citep{ven12}.

The far-infrared continuum, measured from the line free channels in
the spectrum, is detected with a flux density of
$f_c=1.92\pm0.14$\,mJy. The continuum is also not resolved. The
rest-frame \cii\ equivalent width (EW) is 0.43$\pm$0.08\,$\mu$m, which
is a factor $\sim$2 below the median \cii\ EW of starburst galaxies
\citep[which have median
  EW$_\mathrm{[CII]}=1.0$\,$\mu$m,][]{sar14}. The luminosity of the
far-infrared emission depends on the assumed model for the dust
emission. The modified black body ($T_d=47$\,K and $\beta=1.6$) gives
\lfir\,$=(4.5\pm0.3)\times10^{12}$\,\lsun, while scaling the Arp220
and M82 templates to the observed continuum flux density results in a
far-infrared luminosity of $(2.5\pm0.2)\times10^{12}$\,\lsun\ and
$(2.9\pm0.2)\times10^{12}$\,\lsun\ respectively. We therefore estimate
that \lfir\ is in the range $(2.4-4.9)\times10^{12}$\,\lsun. The total
infrared luminosity is calculated to be
$(4.0\pm0.3)\times10^{12}$\,\lsun, $(6.3\pm0.5)\times10^{12}$\,\lsun,
and $(6.4\pm0.5)\times10^{12}$\,\lsun\ for the Arp220 template, the
M82 template, and the modified black body, giving a range of
\ltir\,$=(3.8-6.9)\times10^{12}$\,\lsun. Assuming the total
far-infrared emission is powered by star formation, this results in a
SFR\,$=555-1020$\,\msunyr. Applying the relation between \lcii\ and
SFR gives a lower SFR of SFR\,=\,$270_{-170}^{+410}$\,\msunyr.
Combined with the SFR derived from the TIR luminosity, our best
estimation of the SFR in this quasar host is 100--1020\,\msunyr. The
dust mass is estimated to be in the range
$(2.7-15.5)\times10^8$\,\msun.

\subsection{J0109--3047}
\label{sec:j0109res}

The quasar J0109--3047 has the faintest absolute UV magnitude
\citep[$M_\mathrm{UV}=-25.5$,][]{ven13} of our sample. From the
\mgii\ line a redshift of $z=6.747$ was derived \citep{der14}. In the
ALMA data the \cii\ emission line is clearly detected with a peak
$\mathrm{S/N}\approx\,11$ (Fig.~\ref{fig:maps}), but at a redshift of
$z_\mathrm{[CII]}=6.7909\pm0.0004$, which is
1690$\pm$232\,\kms\ redward of the expected redshift based on the
\mgii\ line. This is a significant offset, and we will discuss the
shifts between the \mgii\ and \cii\ lines in Section
\ref{sec:ciimgii}.

The \cii\ line in the spectrum extracted from the brightest pixel in
our smoothed data cube (Fig.\ \ref{fig:spectra}) has a peak flux of
$f_p=5.6\pm0.5$\,mJy\,beam$^{-1}$ and a FWHM of 340$\pm$36\,\kms. The
line emission is marginally resolved (Fig.~\ref{fig:velmaps}) in the
0\farcs70$\times$0\farcs45 beam with a deconvolved size of
0.43$\pm$0.10 arcsec $\times$ 0.39$\pm$0.15 arcsec. At a redshift of
$z=6.79$ this corresponds to a size of
(2.3$\pm$0.5)$\times$(2.1$\pm$0.8) kpc$^2$. The integrated line flux
is $F_\mathrm{[CII]}=2.04\pm0.20$\,\jykms, and the luminosity is
\lcii\,$=(2.4\pm0.2)\times10^9$\,\lsun, similar to the
\cii\ luminosity of J2348--3054.

The continuum is significantly fainter than that of the other two
quasars, with a measured flux density of $f_c=0.56\pm0.11$\,mJy. The
source is not resolved in the continuum map
(Fig.~\ref{fig:maps}). Given the moderate S/N\,$=7.2$ of the continuum
emission we cannot exclude that the size of the line and continuum
emission are similar. The equivalent width of the \cii\ line is
EW\,$=1.90\pm0.42$\,$\mu$m, which is higher than the \cii\ EW found
for local starburst galaxies \citep{sar14}.

Based on the three different models for the shape of the far-infrared
emission and taking into account the uncertainty in the measured flux
density, our best estimate of the far-infrared luminosity is
\lfir\,$=(0.6-1.5)\times10^{12}$\,\lsun, the total infrared luminosity
is in the range \ltir\,$=(0.9-2.2)\times10^{12}$\,\lsun, and the total
mass of dust is $M_d=(0.7-4.9)\times10^8$\,\msun. From the infrared
luminosity we derive a star formation rate of
SFR\,$=140-325$\,\msunyr. The strength of the \cii\ line results in a
similar SFR of SFR$_\mathrm{[CII]}=355_{-215}^{+540}$\,\msunyr.

\subsection{J0305--3150}
\label{sec:j0305res}

J0305--3150 with $z_\mathrm{MgII}=6.605$ is the brightest of the three quasars
with an absolute UV magnitude of $M_\mathrm{UV}=-26.0$ \citep{ven13}. In the
ALMA data both the \cii\ emission line and the far-infrared continuum are
detected at high significance (S/N\,$>25$, see Fig.~\ref{fig:maps}). The
\cii\ line gives a redshift of the quasar host of
$z_\mathrm{[CII]}=6.6145\pm0.0001$ (Fig.\ \ref{fig:spectra}), which is
slightly higher (374$\pm$79\,\kms) than that of the \mgii\ line. 

The peak flux density of the \cii\ line in Fig.\ \ref{fig:spectra} is
$f_p=12.7\pm0.5$\,mJy\,beam$^{-1}$. The line width is
FWHM\,$=255\pm12$\,\kms. The line emission is resolved with a
deconvolved size of (0.60$\pm$0.03)$\times$(0.42$\pm$0.04) arcsec$^2$,
which corresponds to (3.3$\pm$0.2)$\times$(2.3$\pm$0.2) kpc$^2$. The
integrated line flux of $F_\mathrm{[CII]}=3.44\pm0.15$\,\jykms\ is the
highest of the three quasars and the \cii\ luminosity of
\lcii\,$=(3.9\pm0.2)\times10^9$\,\lsun\ is similar to that of the
$z=6.42$ quasar J1148+5251
\citep[\lcii\,$=(4.1\pm0.3)\times10^9$\,\lsun;][]{mai05,wal09b},
making it one of the brightest \cii\ emitters at $z>6$. If the
\cii\ emission traces star formation activity, then the SFR is
SFR$_\mathrm{[CII]}=630_{-380}^{+955}$\,\msunyr. Taking the diameter
of the line emitting region as 1.5$\times$ the deconvolved size
\citep[e.g.,][]{wan13}, then we derive a source area of $\pi\,(0.75
a_\mathrm{maj})\times(0.75 a_\mathrm{min})=14\pm2$\,kpc$^2$ with
$a_\mathrm{maj}$ and $a_\mathrm{min}$ the deconvolved major and minor
axis FWHM of the line emitting region (see
Table~\ref{tab:firprop}). We determine a \cii\ surface density of
$\Sigma_\mathrm{[CII]}=(2.9\pm0.4)\times10^8$\,\lsun\,kpc$^{-2}$ or
$\Sigma_\mathrm{[CII]}=(1.1\pm0.1)\times10^{42}$\,erg\,s$^{-1}$\,kpc$^{-2}$.
Applying the relation between \cii\ surface density and star formation
rate surface density from \citet{her15} and taking into account that
galaxies with \ltir\,$>10^{11}$\,\lsun\ have a normalization that is a
factor $\sim$2 higher, we derive a SFR surface
$\Sigma_\mathrm{SFR}\sim25$\,\msunyr\,kpc$^{-2}$ and a total
SFR$_\mathrm{[CII]}\sim335$\,\msunyr, approximately a factor two lower
than our other \cii\ SFR estimates.

\begin{deluxetable*}{ccccccc}
\tablecaption{Additional Continuum Sources in the Quasar Fields. 
\label{tab:spurcont}}
\tablewidth{0pt} 
\tablehead{\colhead{Field} & \colhead{R.A. (J2000)} &
  \colhead{Decl. (J2000)} & \colhead{Flux density} & \colhead{S/N} &
  \colhead{$J_\mathrm{AB}$} & \colhead{$K_\mathrm{s,AB}$}}
\startdata
J2348--3054 & 23$^h$48$^m$32$^s\!\!$.92 &
--30$^\circ$54\arcmin06\farcs52 & 0.65$\pm$0.06 & 10.6 &
$>$22.3\tablenotemark{a} & $>$21.5\tablenotemark{a} \\ 
J0305--3150 & 03$^h$05$^m$16$^s\!\!$.37 &
--31$^\circ$50\arcmin54\farcs95 & 0.21$\pm$0.04 & 5.1 &
$>$22.1\tablenotemark{a} & $>$21.6\tablenotemark{a} \\ 
J0305--3150 & 03$^h$05$^m$17$^s\!\!$.11 &
--31$^\circ$50\arcmin52\farcs10 & 0.20$\pm$0.04 & 4.9 &
$>$22.1\tablenotemark{a} & $>$21.6\tablenotemark{a} 
\enddata
\tablenotetext{a}{3$\sigma$ magnitude limits}
\end{deluxetable*}

In the map of the continuum emission (Fig.\ \ref{fig:maps}) we detect
the quasar with a S/N\,$=62$. The detection is at such high
significance that we can even constrain the slope of the continuum
emission of this source (see the discussion in
Section~\ref{sec:j0305cont}). The continuum emission is also resolved
and the deconvolved size of the source is
(0.40$\pm$0.02)$\times$(0.29$\pm$0.02)\,arcsec$^2$, or
(2.2$\pm$0.1)$\times$(1.6$\pm$0.1)\,kpc$^2$. The object appears thus
more extended in the line emission. This could indicate the presence
of an additional component in the \cii\ emission or dust heating by
the central active galactic nucleus (AGN). We will discuss this
further in Sections~\ref{sec:ciifir} and \ref{sec:correlations}.

The continuum flux density measured in the spectrum
(Fig.~\ref{fig:spectra}) is $f_c=3.29\pm0.10$\,mJy, making this one of
the brightest $z>5.5$ quasars observed around 250\,GHz
\citep[e.g.,][]{wan08b}. The \cii\ equivalent width is
EW$_\mathrm{[CII]}=0.55\pm0.03$\,$\mu$m, a factor $\sim$2 lower than
that of local starbursts and similar to the EW$_\mathrm{[CII]}$
measured in J2348--3054 (Section~\ref{sec:j2348res}). From the
measured continuum flux density we derive
\lfir\,$=(4.0-7.5)\times10^{12}$\,\lsun\ and
\ltir\,$=(6.3-10.6)\times10^{12}$\,\lsun\ (but see the discussion in
Section~\ref{sec:cmb}). The total infrared luminosity results in an
upper limit on the SFR of SFR\,$=940-1580$\,\msunyr. However, as will
be discussed in Sections~\ref{sec:j0305cont} and \ref{sec:cmb}, the
TIR luminosity in this quasar host might be overestimated. From the
continuum slope we measured a dust temperature of 30\,K, resulting in
a lower \ltir\ and implying a SFR$_\mathrm{TIR}=545$\,\msunyr, similar
to the \cii\ derived SFR. Finally, we estimate that the dust mass in
this quasar host is in the range $M_d=(4.5-24)\times10^8$\,\msun.

\subsection{Other sources in the field}
\label{sec:companions}

We searched for other sources in the field of the quasars. We searched the
data cubes for emission line sources and the continuum images for continuum
sources.

\subsubsection{Continuum Sources in the Field}
\label{sec:continuumsources}

The continuum images have rms values of 61\,$\mu$Jy\,beam$^{-1}$,
67\,$\mu$Jy\,beam$^{-1}$, and 42\,$\mu$Jy\,beam$^{-1}$ for
J2348--3054, J0109--3047, and J0305--3150, respectively. The largest,
negative noise peaks in the images have a S/N\,$= -4.3$. In the
following we assume that sources with a S/N\,$>4.5$ are real, and not
due to noise fluctuations. In the three quasar fields we discovered
three objects with a peak flux density above a S/N $>4.5$. The
coordinates and flux densities are listed in Table
\ref{tab:spurcont}. We verified that these objects are not artifacts
from the central quasar left over after cleaning. We have checked the
NASA/IPAC Extragalactic Database
(NED)\footnote{\url{http://ned.ipac.caltech.edu}} and none of the
sources had a counterparts in the database. Also, no associated
near-infrared sources were found in the VIKING images down to
$J_\mathrm{AB}>22.1$ and $K_\mathrm{s,AB}>21.5$ (see e.g.,
\citealt{ven13} for details about the VIKING survey).

The Half Power Beam Width (HPBW) of the ALMA 12\,m antennae is $1.13
\lambda /D$\footnote{Remijan et al.\ 2015, ALMA Cycle 3 Technical
  Handbook Version 1.0}. For our setup this translates to a HPBW of
$\sim$24\arcsec--25\arcsec\ (0\farcm40--0\farcm42). The circular area
of the sky given by the HPBW in which we searched for mm sources is
approximately 0.13--0.14\,arcmin$^2$ per field. Based on our number
counts, we derive a source density at
$\lambda_\mathrm{obs}\sim1.3$\,mm of $\sim$2.5\,arcmin$^{-2}$
($\sim$$0.9\times10^4$\,deg$^{-2}$) for sources above a flux density
$f_\nu\gtrsim0.3$\,mJy, the 4.5$\sigma$ limit in the field with the
shallowest continuum data. The source density increases to
$\sim$16\,arcmin$^{-2}$ ($\sim$$6\times10^4$\,deg$^{-2}$) for the
faintest source we can detect ($f_\nu\gtrsim0.2$\,mJy). These counts
are consistent with the mm number counts derived by \citet{car15} and
model predictions by \citet{shi12} and \citet{cai13}. This implies
that the quasars are not located in substantial overdensities of
dust-obscured, highly star-forming galaxies, at least not in the
immediate vicinities of $\sim$12\arcsec\ radius ($\sim$65\,kpc).
 
\subsubsection{Additional Emission Line Objects}
\label{sec:linesources}

We also searched the data cubes for emission line objects in the field. Each
data cube covers roughly 7\,GHz in frequency with a gap between the
bandpasses. We binned the data with various widths, and subsequently searched
for sources with significant (S/N\,$>5$) emission. We did not find any sources
besides the quasar hosts in cubes with channels widths of 75, 125, and
250\,\kms.

We can use this result to set upper limits on the volume density of
\cii\ emission line galaxies at $6.6<z<6.9$. If we assume that
\cii\ emission line galaxies have line width of FWHM\,$=100$\,\kms,
then our 5\,$\sigma$ limits are 0.26\,Jy\,\kms\ for the J2348--3054
and J0109--3047 fields, and 0.16\,Jy\,\kms\ for the J0305--3150
field. The corresponding luminosity limits for the fields are
\lcii\,$>3.1\times10^8$\,\lsun, and \lcii\,$>1.8\times10^8$\,\lsun,
respectively. For objects with an emission line of 50\,\kms\ this
limit is $\sqrt{2}$ lower. The redshift range probed by our
observations is roughly $\Delta z = 0.21-0.24$, and the total volume
we probe is 205\,comoving Mpc$^3$. Using Poisson statistics
\citep{geh86} we can set 1 sigma upper limits on the space density of
\cii\ emitters at $6.6<z<6.9$ of
$\rho($\lcii\,$>3.1\times10^8$\,\lsun$)<9\times10^{-3}$\,Mpc$^{-3}$
and
$\rho($\lcii\,$>1.8\times10^8$\,\lsun$)<3\times10^{-2}$\,Mpc$^{-3}$.

\section{DISCUSSION}
\label{sec:discussion}

\subsection{\cii\ Line to FIR Luminosity ratio}
\label{sec:ciifir}

\begin{figure}
\includegraphics[width=\columnwidth]{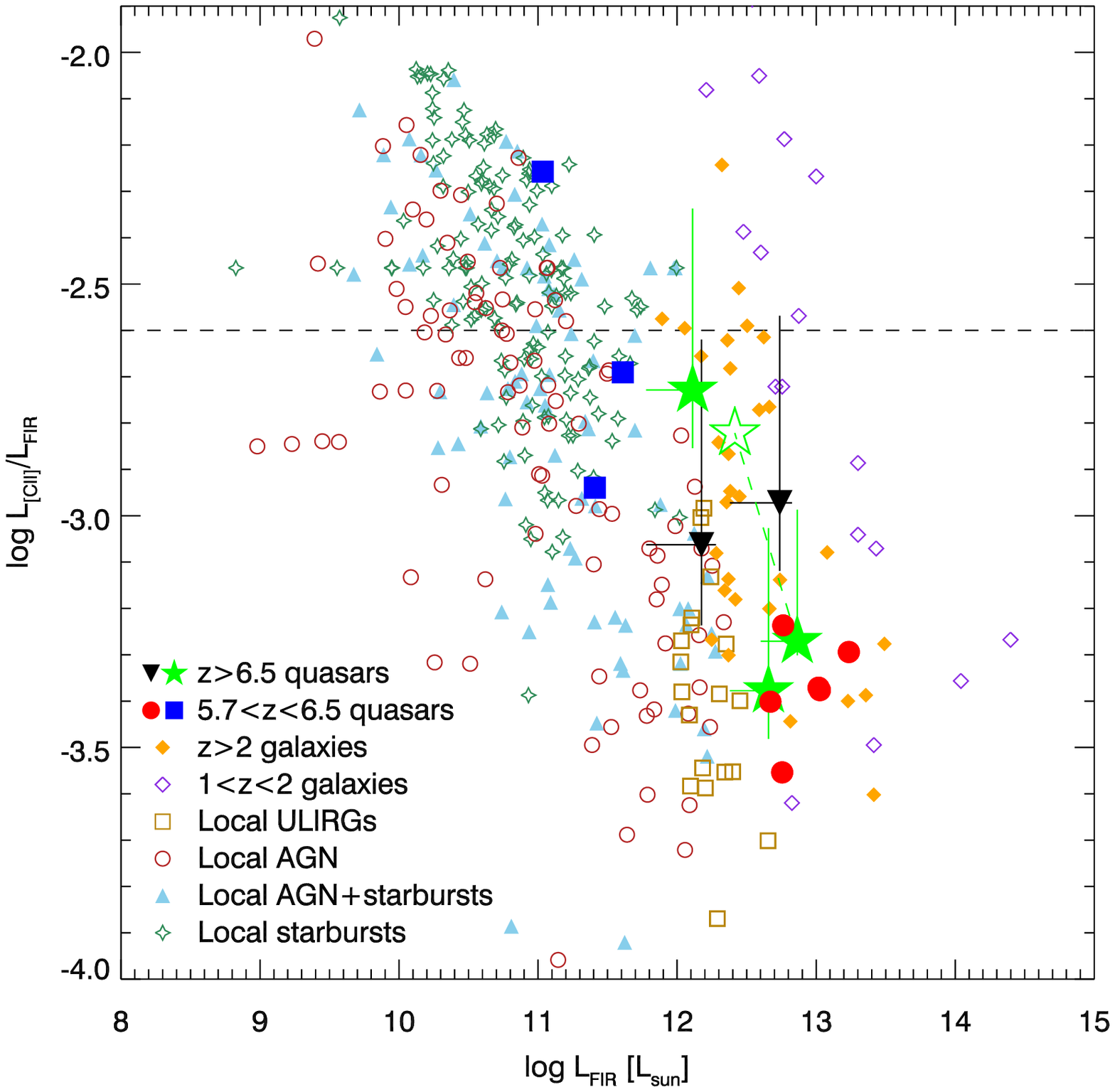}
\caption{Ratio of \cii\ luminosity over FIR luminosity as function of
  FIR luminosity. Plotted are values for local star-forming and
  starburst galaxies (open stars), local AGN (open circles), and local
  composite galaxies \citep[filled triangles, all
    from][]{mal01,sar12,sar14,dia13}. We further plot local ultra
  luminous infrared galaxies \citep[ULIRGS, open squares,][]{far13},
  galaxies at $1<z<2$ \citep[open diamonds,][]{sta10,bri15} and
  $2<z\lesssim6$ galaxies \citep[filled
    diamonds,][]{mai09,ivi10,wag10,deb11,cox11,rie14,gul15}. Data
  points from lensed objects were corrected for the
  magnification. Quasar host galaxies at $5.7<z<6.5$
  \citep{mai05,wal09b,wan13,wil13,wil15} are indicated with filled
  circles and squares. Finally, the filled stars and filled
  upside-down triangles present quasar hosts at $z>6.5$. The triangles
  indicate the values found for P036+03 at $z=6.541$ \citep{ban15b}
  and J1120+0641 at $z=7.084$ \citep{ven12}. The values for the three
  VIKING quasars at $6.6<z<6.9$ presented in this work are plotted as
  green stars. The plotted values are calculated assuming that the
  dust has a temperature of 47\,K and an emissivity of $\beta=1.6$, as
  is typically assumed for $z\sim6$ quasar hosts
  \citep[e.g.][]{wan13,wil15}. The big open star indicates the
  \cii--to--FIR luminosity ratio for J0305--3150 with a dust
  temperature of $T_d\sim30$\,K, which was measured by fitting the
  continuum slope in the quasar host (see Section~\ref{sec:cmb}). The
  \lcii/\lfir\ ratio found for $z>6.5$ quasars span a range of values,
  similar to the $5.7<z<6.5$ quasar hosts. The error bars take into
  account both the uncertainties in the measurements and the
  uncertainties in the properties of the dust (see
  Section~\ref{sec:luminosities}), showing that the unknown shape of
  the dust emission results in a highly uncertain \cii--to--FIR
  luminosity ratio in $z>6.5$ quasar host
  galaxies. \label{fig:lciilfir}}
\end{figure}

\begin{figure}
\includegraphics[width=\columnwidth]{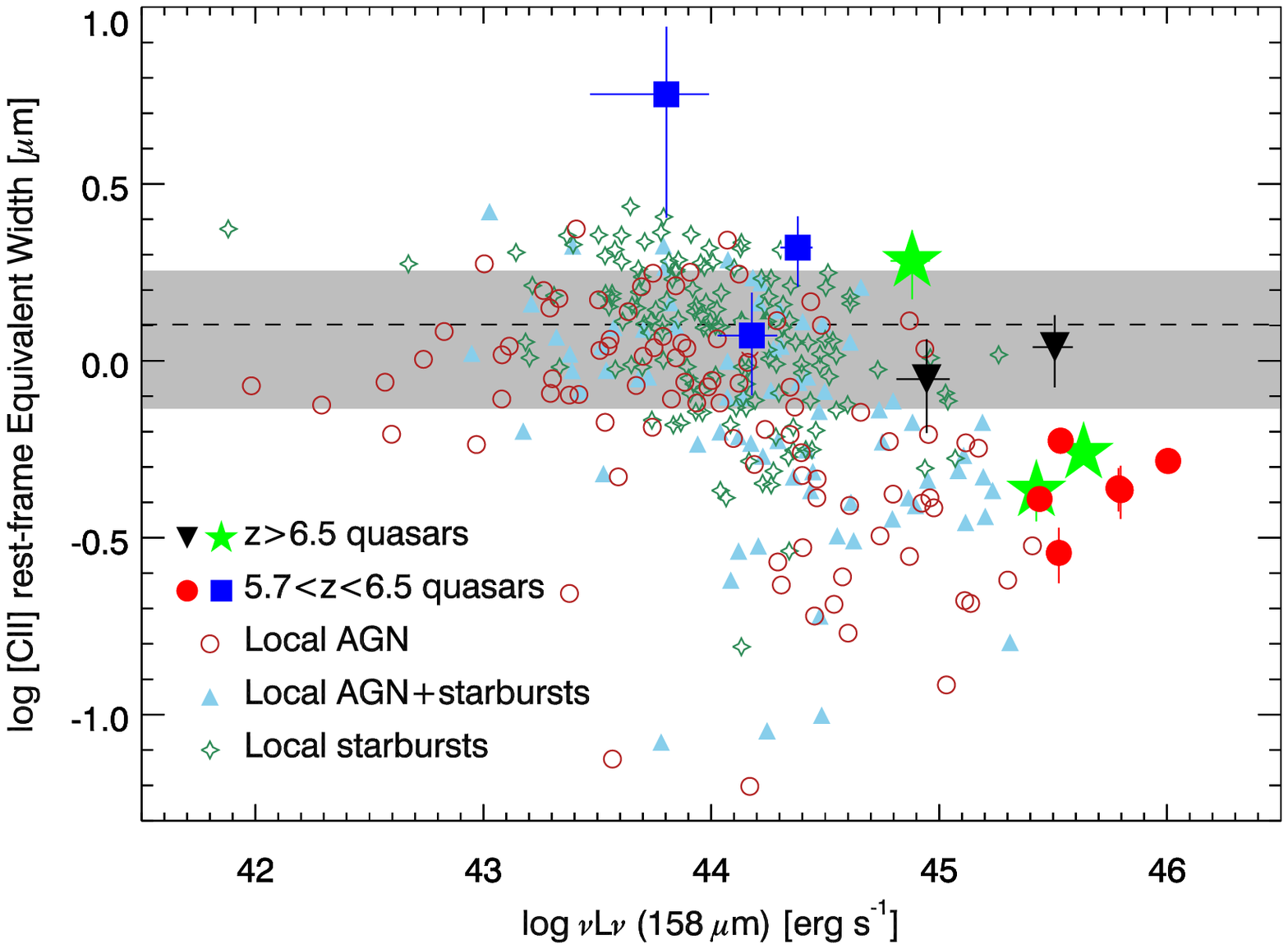}
\caption{Rest-frame equivalent width (EW) of the \cii\ line
  vs. monochromatic luminosity at a wavelength of 158\,$\mu$m
  (rest-frame). We plotted values for local systems (starburst
  galaxies as open stars, AGN as open circles and composite galaxies
  as filled triangles) from \citet{dia13} and \citet{sar14}. We marked
  the mean and 1\,sigma range found for local starburst galaxies with
  a dashed line and gray region. As in Fig.~\ref{fig:lciilfir}, the
  blue squares and red circles indicate the values found for quasars
  at $5.7<z<6.5$, while the green filled stars and black, upside down
  triangles are the VIKING quasars and two other $z>6.5$ quasars. The
  high redshift quasars have \cii\ EWs within a factor of $\sim$5 of
  the mean of that of local starbursts.\label{fig:ciiew}}
\end{figure}

\begin{figure*}
\includegraphics[width=\textwidth]{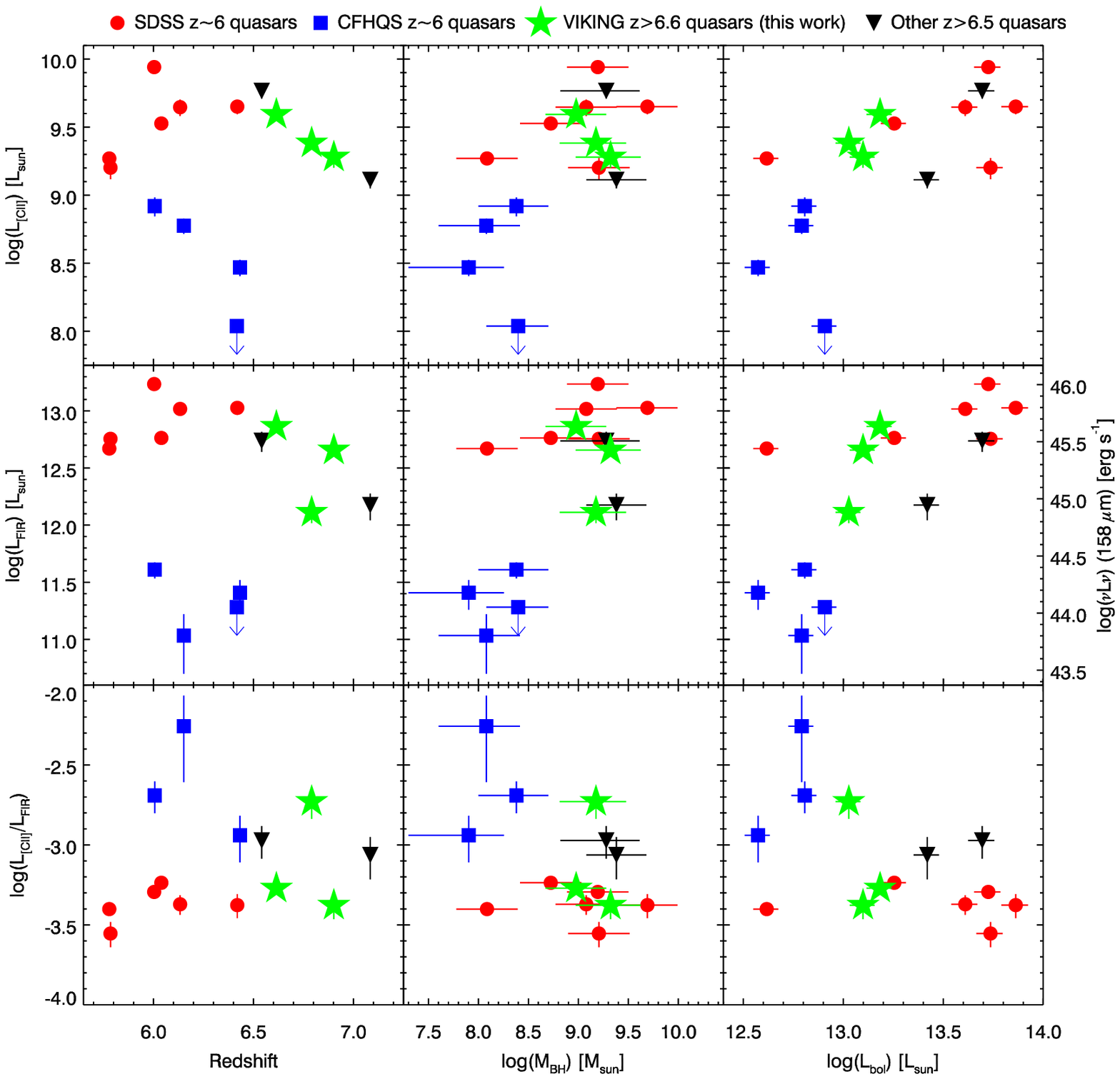}
\caption{Host galaxy properties measured in the far-infrared (\lcii,
  \lfir, and \lcii/\lfir) plotted against the characteristics of the
  quasar ($M_\mathrm{BH}$ and $L_\mathrm{bol}$) and redshift. The
  bolometric luminosity of the quasars was calculated by applying the
  bolometric correction to the absolute magnitudes at 1450\,\AA\ in
  the rest-frame ($M_{1450}$) from \citet{run12}, with the absolute
  magnitudes taken from the literature \citep{cal14,ven13,ven15a}. The
  red circles represent the properties of $z\sim6$ SDSS quasars
  observed by \citet{mai05,wan13} in the far-infrared. The blue
  squares are CFHQS quasars at $6.0<z<6.4$ published by
  \citet{wil10b,wil13,wil15}. The black triangles are two $z>6.5$
  quasars previously observed in the far-infrared (published by
  \citealt{ven12} and \citealt{ban15b}). The black hole masses of
  these two quasars were taken from \citet{der14} and
  \citet{ven15a}. The green stars are the $6.6<z<6.9$ quasars
  presented in this paper. The black hole masses of these
  quasars were published in \citet{der14}. For consistency, \lfir\ has
  been computed for all sources assuming $T_d=47$\,K and
  $\beta=1.6$. \label{fig:correlations}}
\end{figure*}

We first study the common \cii\--to--FIR luminosity ratios of
our sample and compare these to the ratios found for lower redshift
objects. Previous studies of $z\sim6$ quasar hosts have shown a range
of results. For example, the SDSS quasars studied by \citet{wan13} and
\citet{mai05} all have low \lcii/\lfir\ ratios
(log(\lcii/\lfir)\,$\sim-3.5$), suggesting that quasar host galaxies
are similar to local ULIRGs \citep[e.g.,][]{far13}. On the other hand,
\citet{wil13,wil15} looked at $z\sim6$ quasars from the CFHQS, which
are fainter and have lower ($<10^9$\,\msun) black hole masses, and
found that \lfir\ for these quasars is lower and the
\lcii/\lfir\ ratio consistent with local star-forming galaxies
($<$log(\lcii/\lfir)$>$\,$=-2.5$; e.g., \citealt{dia13}). A difference
between these studies could be that \citet{wan13} targeted quasars
that were known to be bright in the far-infrared from continuum
studies. Indeed, mm observations of two quasar hosts at $z>6.5$,
P036+03 at $z=6.5$ \citep{ban15b} and J1120+0641 at $z=7.1$
\citep{ven12}, revealed bright far-infrared continua
(\lfir\,$>10^{12}$\,\lsun), but \lcii/\lfir\ ratios close to that
star-forming galaxies, log(\lcii/\lfir)\,$=-3$.

The newly detected quasar hosts presented in this paper show a range
of properties (Fig.~\ref{fig:lciilfir}). While two of the sources,
J0305--3150 and J2348--3054, have ratios similar to the quasars
studied by \citet{wan13}, J0109--3047 has a ratio consistent with
star-forming galaxies. The range of characteristics of $z>6.5$ quasar
hosts is quite similar to that of $z>2$ (ultra-)luminous infrared
galaxies, roughly following the correlation between \lfir\ and
\lcii/\lfir\ discussed in \citet{wil15}. A possible explanation for
the decreasing \lcii/\lfir\ ratio as function of increasing
\lfir\ could be that in the $z\gtrsim6$ quasar hosts studied here and
by, e.g., \citet{wan13} and \citet{wil15}, at least a fraction of the
FIR luminosity is due dust heating by the central AGN. Alternatively,
the strong X-ray radiation from the central source could affect the
C$^+$ abundance, reducing the \cii\ luminosity
\citep[e.g.,][]{lan15}. 

An issue with the \cii--to--FIR luminosity ratio in the $z>5.7$
quasar hosts is the unknown shape of the far-infrared dust continuum,
resulting in a highly uncertain estimate of the FIR luminosity (see
Section~\ref{sec:luminosities}). By analyzing the spectral energy
distribution of far-infrared bright
($f_\mathrm{obs}(1.2\mathrm{mm})>1$\,mJy) quasars at $z>5$,
\citet{lei13} found dust temperature in the range $T_d=40-60$\,K. In
the literature, a dust temperature of 47\,K is regularly assumed,
even for $z\sim6$ quasar hosts with relative weak
($f_\mathrm{obs}(1.2\mathrm{mm})\lesssim0.2$\,mJy) far-infrared
continua \citep[e.g.,][]{wil15}. If the dust temperature varies
significantly among different quasar host galaxies (see
Section~\ref{sec:j0305cont} for an example), the spread in
\lcii/\lfir\ ratio could be larger than shown in
Fig.~\ref{fig:lciilfir}.

A more direct measurement of the relative strength of the \cii\ line
with respect to the underlying continuum can be obtained from our data
by dividing the line flux by the continuum flux density: the
\cii\ equivalent width (EW). The advantage of calculating the
\cii\ equivalent width over the \lcii/\lfir\ ratio is that it does not
depend on the characteristics of the dust continuum emission. For our
quasar hosts, we obtained rest-frame \cii\ equivalent widths between
0.43 and 1.9\,$\mu$m (Fig.~\ref{tab:firprop}). These values are within
a factor of $\sim$3 of the mean EW$_\mathrm{[CII]}=1.27$\,$\mu$m found
for local starburst galaxies
\citep[Fig.~\ref{fig:ciiew};][]{dia13,sar14}.

In the next section, we will compare the properties of the
far-infrared emission of the quasar hosts with those of their nuclear
source.

\subsection{Correlations Between UV and FIR Properties}
\label{sec:correlations}

In Fig.~\ref{fig:correlations} we compare the FIR properties (\lcii,
\lfir, and \lcii/\lfir) of the $z>5.7$ quasar hosts with the redshift
and the characteristics of the accreting black hole (black hole mass
$M_\mathrm{BH}$ and bolometric luminosity $L_\mathrm{bol}$ of the
central source). The bolometric luminosity $L_\mathrm{bol}$ of the
central AGN was computed by applying a bolometric correction to the
monochromatic luminosity at 1450\,\AA\ in the rest-frame. The
monochromatic luminosities were derived from published absolute
magnitudes \citep{cal14,ven13,ven15a}, which have an assumed
uncertainty of 15\%. We derived the bolometric correction by taking
the data from Table 1 in \citep{run12} and fitting a line of the form:

\begin{equation}
\mathrm{log}\left(\frac{L_\mathrm{bol}}{10^{46}
\,\mathrm{erg\,s}^{-1}}\right)\,=\,a\,+b\,\mathrm{log}\left(\frac{\lambda
L_\lambda}{10^{46}\,\mathrm{erg\,s}^{-1}}\right).
\label{eq:lbol}
\end{equation}

\noindent
We obtain $a=0.459\pm0.017$ and $b=0.911\pm0.022$\footnote{In
  \citet{run12} they fit a line
  $\mathrm{log}(L_\mathrm{iso})\,=\,a\,+b\,\mathrm{log}(\lambda
  L_\lambda)$ with $L_\mathrm{iso} = L_\mathrm{bol} / 0.75$. In this
  form, $a$ is determined where $\mathrm{log}(\lambda L_\lambda)=0$,
  which is far from the range of $\lambda L_\lambda$ that was
  fitted. As a consequence, the uncertainty in $a$ is large (1\,dex)
  and the resulting uncertainty in $L_\mathrm{bol}$ computed using
  their best fit parameters is overestimated.} and use these
parameters to compute $L_\mathrm{bol}$ from the monochromatic
luminosity at $\lambda_\mathrm{rest}=1450$\,\AA.

The black hole masses of the VIKING quasars, estimated from the
width of the \mgii\ line and the strength of the quasar's continuum,
were derived in \citet{der14}. For the other $z\sim6$ quasars, black
hole masses derived from the \mgii\ line were taken from the
literature when available \citep{wil10b,der11,der14,ven15a}. For
objects for which no \mgii\ derived black hole masses are published,
we assumed that the quasars are accreting at the Eddington luminosity
($L_\mathrm{Edd}=1.3\times10^{38} (M_\mathrm{BH}/M_\odot)$\,erg\,s$^{-1}$)
as has been found for $z\sim6$ quasars
\citep[e.g.,][]{wil10b,der11}. To account for the range in Eddington
ratios observed in these quasars, we added an uncertainty of 0.3\,dex
in quadrature to the uncertainty in the bolometric luminosity. For
easy comparison we computed \lfir\ assuming a modified black body with
$T_d=47$\,K and $\beta=1.6$ for all sources (however, see
Section~\ref{sec:j0305cont} for a discussion on this assumption).

From the parameters plotted against each other in
Fig.~\ref{fig:correlations}, four correlate strongly (defined by us as
having a Pearson's $r$ of $|r|>0.5$): $M_\mathrm{BH}$ with \lfir\ or,
more accurately, with the measured monochromatic luminosity at a
rest-frame wavelength of 158\,$\mu$m, $\nu L_{\nu,158\,\mu\mathrm{m}}$, 
$M_\mathrm{BH}$ with \lcii, and $L_\mathrm{bol}$ with \lfir\ (or $\nu
L_{\nu,158\,\mu\mathrm{m}}$) and \lcii. Since all quasars plotted in the
figure are either accreting close to the Eddington limit or explicitly
assumed to accrete at Eddington, the correlation between
$M_\mathrm{BH}$ and \lfir\ (\lcii) could be due to the correlations
between $L_\mathrm{bol}$ and \lfir\ (\lcii). Although the strong
correlation ($r=0.72$) between $L_\mathrm{bol}$ and the far-infrared
continuum luminosity could suggest that part of the FIR emission is
coming from dust heated by the AGN, a similarly strong correlation
($r=0.67$, or $r=0.72$ excluding the undetected quasar) can be seen
between $L_\mathrm{bol}$ and \lcii. Furthermore, the
\lcii/\lfir\ ratio correlates only weakly ($r=-0.38$) with
$L_\mathrm{bol}$, with quasars occupying a range in \lcii/\lfir\ at
both low and high $L_\mathrm{bol}$. By fitting a line through the
\lcii-\lbol\ and $\nu L_{\nu,158\,\mu\mathrm{m}}$-\lbol\ data, we found the
following relations:

\begin{eqnarray}
  \nu\,L_\nu (158\,\mu\mathrm{m})=10^{44.94\pm0.15} \left(\frac{L_\mathrm{bol}}{10^{13} L_\odot}\right)^{1.09\pm0.30} L_\odot, \\
  L_\mathrm{[CII]}=10^{9.16\pm0.09} \left(\frac{L_\mathrm{bol}}{10^{13} L_\odot}\right)^{0.68\pm0.18} L_\odot.
\end{eqnarray}

As mentioned in Section~\ref{sec:ciifir}, a possible explanation for
low \lcii/\lfir\ ratios (the ``\cii\ deficit'') in quasar hosts is
that the strong X-ray radiation from the central AGN reduces the $C^+$
abundance and hence suppresses the \cii\ emission \citep{lan15}. The
positive slope in the \lcii-\lbol\ relation indicates that, in the
quasars studied here, this scenario is not the explanation for the
\cii\ deficit. The positive correlation between the far-infrared
continuum and \lbol\ might indicate that a fraction of \lfir\ could be
due to dust heated by the AGN. This is supported by the size
estimations of J0305--3150 (Table~\ref{tab:firprop}) and of bright
quasar hosts at $z\sim6$ \citep{wan13}: the region emitting the
continuum radiation seems to be smaller than the \cii\ emitting
region. However, this does not explain why the \cii\ luminosity also
correlations with the luminosity of the quasar.

\begin{figure*}
\includegraphics[width=\textwidth]{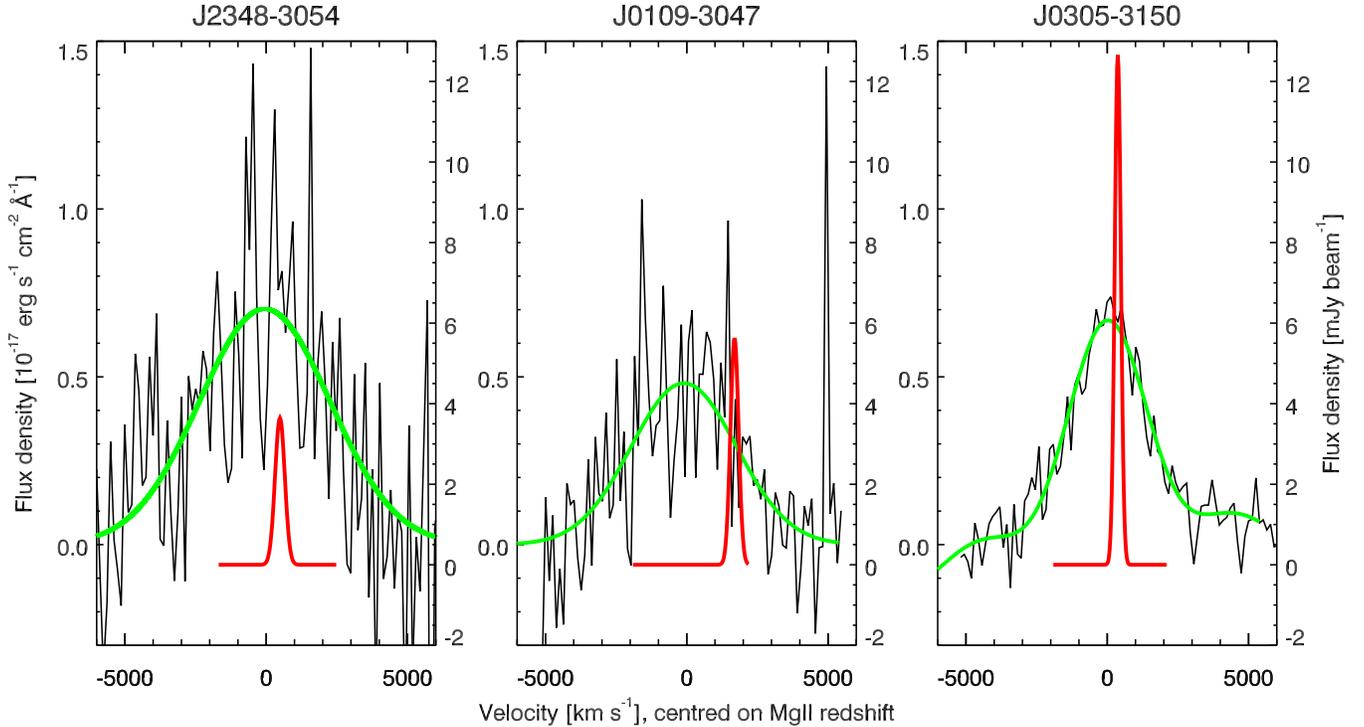}
\caption{Continuum subtracted near-infrared spectra around the
  \mgii\ emission line of the three quasars (black thin lines) with
  the model of the \mgii\ emission line overplotted in green (taken
  from Fig.\ 4 of \citealt{der14}). In red the Gaussian fits to the
  \cii\ emission line (see Fig.\ \ref{fig:spectra}) are shown. The
  left y-axis gives the flux units of the \mgii\ line, whereas the
  right y-axis gives the units of the \cii\ line emission. On the
  x-axis we plot the velocity with respect to the \mgii\ redshift. In
  particular in the case of J0109--3047, the peak of the \mgii\ and
  \cii\ lines show significant offsets from each other (see discussion
  in Section~\ref{sec:ciimgii}). \label{fig:mgiicii}}
\end{figure*}

A likely scenario that produces a positive correlation between both
\lcii-\lbol\ and \lfir-\lbol\ is that a large reservoir of gas is
available to both feed the black hole and to form stars. This is in
rough agreement with the results of, for example, \citet{lei14} and
\citet{bar15} who found that the FIR flux density in $z=5-7$ quasars
measured around 160\,$\mu$m is dominated by cold dust emission powered
by star formation.

\subsection{\cii\ vs \mgii\ Redshifts}
\label{sec:ciimgii}

As mentioned in Section \ref{sec:luminosities}, the \cii\ emission
line did not coincide with the redshift expected from the
\mgii\ line. In one case, J0109--3047, the shift is
$\sim$1700\,\kms. Since the \mgii\ line is originating from the broad
line region, its line width is much larger than that of the
\cii\ line. One question is whether the shifts could be caused by the
uncertainty in the determination of the centre of the \mgii\ line. In
Fig.~\ref{fig:mgiicii} we plotted the (continuum-subtracted) spectra
of the quasars around the \mgii\ line with the best-fit model for the
emission line on top \citep[see][for the details]{der14}. In the case
of J2348--3054 and J0305--3150, the \cii\ line is located relatively
close ($<$500\,\kms) to the peak of the \mgii\ line that in these
quasars has a FWHM of 3200--5500\,\kms\ \citep{der14}. On the other
hand, the \cii\ line in J0109--3047 is clearly at a higher redshift
than the peak of the \mgii\ emission.

\begin{figure}
\includegraphics[width=\columnwidth]{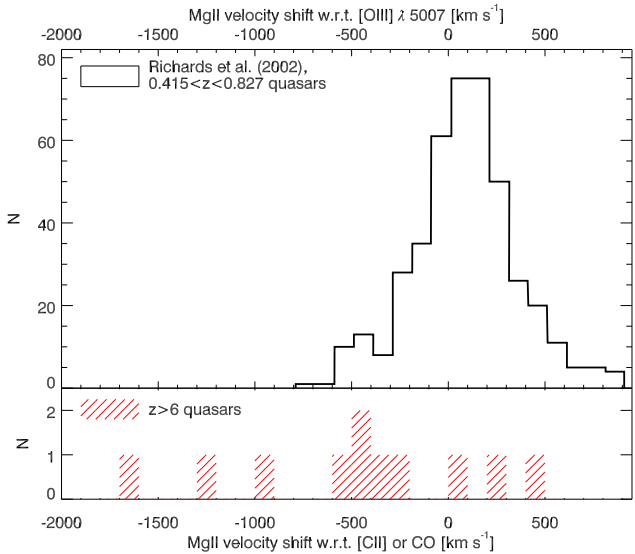}
\caption{{\it Top:} Histogram showing the distribution of the velocity
  shift of the \mgii\ emission line center with respect to that of the
  [\ion{O}{3}] $\lambda$ 5007 line of 417 quasars at $0.415<z<0.827$
  \citep[from][]{ric02b}. Negative velocities indicate a blueshift of
  the \mgii\ line compared to the [\ion{O}{3}] line. {\it Bottom:} In
  the red hashed histogram we plot the velocity shift between the
  redshift determined from the \mgii\ line and that of the quasar host
  galaxy traced by \cii\ or CO emission of $z>6$ quasars. While in low
  redshift quasar spectra the \mgii\ line is redshifted, on average,
  by $97\pm269$\,\kms\ \citep{ric02b}, at $z>6$ the \mgii\ line is
  predominantly blueshifted with respect to the host galaxy redshift
  with a mean and standard deviation of
  $-480\pm630$\,\kms. \label{fig:velshifts}}
\end{figure}

These shifts between the \mgii\ line and the host galaxy redshift as
traced by the \cii\ line are unexpected, as it has been shown that the
\mgii\ line is a good tracer of the systemic redshift at lower
redshifts \citep[e.g.,][]{ric02b,hew10}. For example, \citet{ric02b}
studied the SDSS spectra of 417 quasars at $0.415<z<0.827$ that
contained both the \mgii\ line and the narrow emission line
[\ion{O}{3}] $\lambda$ 5007 line. They found that the \mgii\ line has
a shift of only $97\pm269$\,\kms\ (see also
Fig.~\ref{fig:velshifts})\footnote{In general, the [\ion{O}{3}]
  $\lambda$ 5007 traces the systemic redshift of quasars very well,
  with an average shift between the [\ion{O}{3}] line and the systemic
  redshift of 40--45\,\kms\ \citep[e.g.,][]{bor05,hew10}. However, in
  some cases the [\ion{O}{3}] line can display large offsets of up to
  400\,\kms, especially in quasars with a high accretion rate
  \citep[e.g.,][]{bor05,bae14}.}.

To examine whether the \mgii\ line in $z\gtrsim6$ quasar spectra
provide a good measure of the systemic redshift, we compiled a list
with all quasars at these redshifts that have a redshift measurement
from both the \mgii\ line and a molecular or atomic line (CO or
\cii). We plot the computed velocity shifts from this sample in
Fig.~\ref{fig:velshifts}. The shifts span a large range from
$+475$\,\kms\ (redshift) to $-$1700\,\kms\ (blueshift). The mean and
the median of this sample are --480 and --467\,\kms, respectively,
with a standard deviation of 630\,\kms. The distribution is not
centered around 0\,\kms: of the 11 $z\gtrsim6$ quasars, 8 have a
blueshifted \mgii\ line with respect to the systemic velocity as
traced by the molecular or atomic lines. This argues against the
scenario in which the shifts are mainly caused by the uncertainties in
determining the center of the broad \mgii\ emission line, although
further studies are needed to investigate possible systematics
affecting $z_\mathrm{MgII}$.

A possible explanation for the large blueshifts is that the broad line
region close to the black hole, where the \mgii\ emission originates,
is pushed outwards by the strong radiation of the quasar.
We investigated whether there is a correlation between the velocity
offset and the Eddington ratio $L_\mathrm{bol}/L_\mathrm{Edd}$. These
two parameters are only marginally correlated ($r=0.42$). Furthermore,
the trend is opposite to our expectations: the quasars with the
highest accretion show only small blueshifts, while the object with
the largest blueshift, J0109--3047, has a relatively low Eddington
ratio of $L_\mathrm{bol}/L_\mathrm{Edd}\sim0.2$ \citep[see
  also][]{der14}. Also, no correlation was found between the velocity
offset and the bolometric luminosity ($r=0.03$) and between the offset
and the FIR luminosity ($r=-0.02$).

Although our sample is small, the wide distribution of velocity shifts
between $z_\mathrm{MgII}$ and $z_\mathrm{[CII]}$ suggest that caution
should be taken when using the redshift of the \mgii\ line as proxy
for the systemic redshift. For example, FIR lines may be shifted out
of an ALMA bandpass (covering $\sim$2250\,\kms\ at 250\,GHz), leading
to an apparent non-detection.

\subsection{Constraints on J0305--3150}
\label{sec:j0305}

The S/N of the observations of J0305--3150 is high enough that we can
investigate the properties of this particular source in more detail.

\subsubsection{Dynamical Modeling}
\label{sec:dynmodels}

\begin{figure*}
\includegraphics[width=\textwidth]{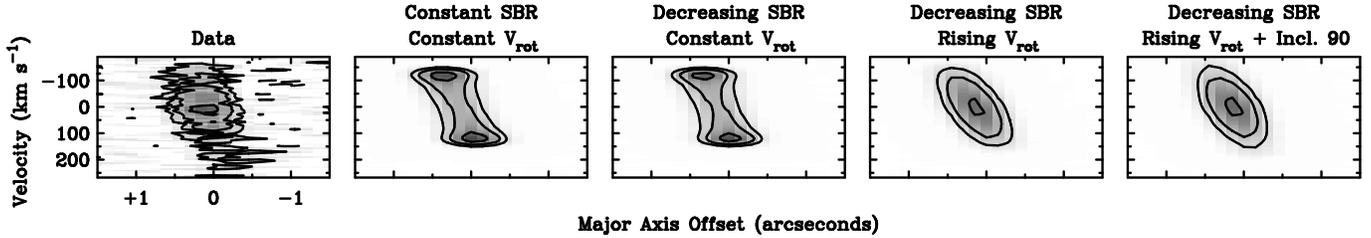}
\caption{Major axis position-velocity diagrams of the data, compared
  with simple theoretical models convolved to our instrumental
  resolution. The models are: a model with both constant surface
  brightness and rotational velocity with radius, a model with a
  constant rotational velocity of 150\,\kms\ with decreasing surface
  brightness, a model with decreasing surface brightness and rising
  rotation curve (from 0 to 150\,\kms\ at 0\farcs1, which corresponds to
    0.55\,kpc at $z=6.6145$), and a model with
  decreasing surface brightness, rising rotation curve (form 0 to
  180\,\kms\ at 0\farcs1), and an inclination of 90$^{\circ}$. All
  other model inclinations are 60$^{\circ}$. Note that a rising
  rotation curve is clearly necessary to provide a match to the data,
  but at the current resolution there is a degeneracy between
  inclinations, and thus the intrinsic rotation
  speed. \label{fig:lvplot}}
\end{figure*}

From Fig.~\ref{fig:velmaps} it is clear that the red and blue side of
the emission line in J0305--3150 are displaced from each other. This
is an indication that ordered motion is present in this quasar host,
and it is possible that the gas is located in a rotating disk. In
contrast to the other two quasar hosts, the \cii\ line in J0305--3150
has been detected at high enough S/N (S/N\,$>25$) to permit modeling
of the gas emission.

Empirical tilted-ring models were created to match the data using the Tilted
Ring Fitting Code \citep[TiRiFiC,][]{joz07}. These models include a single
disk component of constant scale height. Our data clearly rule out models with
a constant surface brightness and a constant rotation velocity (here set to
150\,\kms, see Fig.~\ref{fig:lvplot}). A decreasing surface brightness
distribution allowed for an improved fit to the data, followed by the addition
of a linearly increasing rotation curve (starting at 0\,\kms\ and peaking at
150\,\kms\ at 0\farcs1 (which corresponds to 0.55\,kpc at $z=6.6415$)
before remaining constant for larger radii), which improved the fit
substantially. Although the resolution is poor, from these models it is clear
that we can rule out a flat rotation curve and that the rotation curve is
increasing instead. However, from the final two panels of
Fig.~\ref{fig:lvplot}, it is seen that the inclination cannot be well
constraint at our current resolution. If the gas in this quasar host galaxy is
distributed in a disk, we only observe the rising part of the rotation
curve. In other words, with the current data we cannot independently determine
the inclination angle and the peak velocity and thus, ultimately, the
dynamical mass. We will estimate a dynamical mass from the observed line width
for this and the other two quasar hosts in Section~\ref{sec:mdyn}.

\subsubsection{Additional Emission Components}
\label{sec:j0305wing}

The high S/N of the \cii\ line in J0305--3150 enables us to search for
emission that deviates from the Gaussian fit. From the spectrum in
Fig.~\ref{fig:spectra} we identified highly significant
($\sim$7$\sigma$) excess emission on the red side (low frequency side)
of the Gaussian emission line. The map of this excess emission is
shown in Fig.~\ref{fig:velmaps} as green contours. The flux density of
this excess emission above the Gaussian fit is $1.69\pm0.23$\,mJy. The
location of this emission is significantly offset from the central
line emission by 0\farcs41$\pm$0\farcs04 ($2.3\pm0.2$\,kpc). Similarly
blueshifted emission is not seen on the other side of the
\cii\ line. The origin of this second component is unclear: it could
be an outflow, inflowing gas, or a close companion to the
quasar. Higher S/N and/or higher spatial resolution will help to
distinguish these different cases.

\subsubsection{Continuum Slope}
\label{sec:j0305cont}

\begin{figure}
\includegraphics[width=\columnwidth]{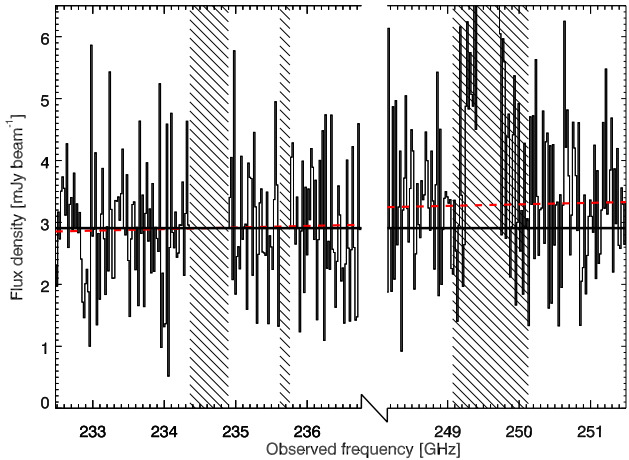}
\caption{Spectrum extracted at the brightest pixel in the data cube of
  J0305--3150 smoothed with a 1\arcsec\ Gaussian, binned over
  20\,MHz. The hashed regions (containing a gap in the frequency
  coverage and the \cii\ line) are not used when computing the
  continuum flux density. The black solid line shows the average
  continuum flux density measured between 232.5\,GHz and
  236.7\,GHz. An increase of the continuum flux density is clearly
  apparent over the frequency range covered by our observations. The
  continuum near the \cii\ line (redshifted to 249.6\,GHz) is
  $\sim$16\% higher. The red, dashed line shows the best fitting
  modified black body. \label{fig:j0305fullspec}}
\end{figure}

The frequency setup of the ALMA observations allows us to measure the
dust continuum in the quasar host around two frequencies that are
roughly 15\,GHz apart. For a source at $z=6.6$, this is $\sim$115\,GHz
in the rest-frame. Over this large frequency range the dust continuum
is not constant. In Fig.~\ref{fig:j0305fullspec} we plot the spectrum
of J0305--3150 in all four bandpasses. The higher frequency data cube
was smoothed with a 1\arcsec\ Gaussian and the extracted spectrum
(between 248.1 and 251.5\,GHz) is identical to the spectrum shown in
Fig.~\ref{fig:spectra}. Because the source is resolved
(Table~\ref{tab:firprop}) and the resolution of the data changes over
the frequency range probed by the observations, we smoothed the lower
frequency data to match the resolution of the (smoothed) higher
frequency data. The average continuum level in the observed frequency
range 232.5--236.7\,GHz is $2.91\pm0.07$\,mJy. If we exclude a region
of $\sim$1.0\,GHz (5$\times$ the FWHM of the \cii\ line) wide around
the \cii\ line, we measure an average continuum level of
$3.29\pm0.10$\,mJy around 250\,GHz. The continuum flux density around
250\,GHz is $0.38\pm0.12$\,mJy higher than around 234\,GHz, a
difference of 3.1$\sigma$. Assuming that this difference is caused by
the shape of the dust continuum emission, we can put constraints the
temperature of the dust. Fitting a modified black body with a fixed
$\beta=1.6$ to the continuum spectrum shown in
Fig.~\ref{fig:j0305fullspec} results in a best-fitting temperature of
$T_d=37_{-7}^{+11}$\,K. We estimated the uncertainty in the
temperature by randomly adding noise to the spectrum and remeasuring
the best-fitting dust temperature 10,000 times. For the 1\,$\sigma$
uncertainties we took the range of temperatures of 68\% of the values
around the median. Since we are in the Rayleigh-Jeans tail of the
modified black body, the uncertainties are asymmetric and
non-Gaussian. The 95\% range (2\,$\sigma$) of the best-fitting dust
temperature is 25--74\,K. Since the dust temperature we derive from
fitting the continuum is only 16\,K above the CMB temperature at this
redshift, $T_\mathrm{CMB}(z=6.61)=20.8$\,K, we consider the effects of
the CMB on the observed dust emission in the next section.

\subsection{Effects of the Cosmic Microwave Background}
\label{sec:cmb}

The effects of the CMB on millimeter observations of high redshift
galaxies are extensively discussed in \citet{dac13}. To summarize, when
the CMB temperature is close to the temperature of the dust in a high
redshift galaxy, there are two competing processes that impact the
observed mm luminosity of the galaxy. Firstly, the CMB supplies an
additional source that heats the dust. The higher dust temperature can
be calculated with the following formula from \citet{dac13}:

\noindent
\begin{equation}
T_d(z)=[(T_d^{z=0})^{4+\beta}+(T_\mathrm{CMB}^{z=0})^{4+\beta}\times((1+z)^{4+\beta}-1)]^{\frac{1}{4+\beta}},
\label{eq:tdz}
\end{equation}

\noindent
where $T_d^{z=0}$ is the dust temperature ignoring heating by the CMB
and $T_\mathrm{CMB}^{z=0}$ is the CMB temperature at $z=0$. The
increase in dust temperature due to heating by the CMB is negligible
for $T_d=47$\,K and $\beta=1.6$ that we assumed in
Section~\ref{sec:results}. If the dust temperature is 30\,K (within
1$\sigma$ of our best-fitting temperature for J0305--3150, see
Section~\ref{sec:j0305cont}), the CMB increases the dust temperature
by $\sim$2\% at $z=6.6$.

The second effect of the CMB on our observations is that it reduces
the detectability of the dust continuum. The fraction of the flux
density that we measure against the CMB is:

\noindent
\begin{equation}
  f_\nu^{\mathrm{obs}}/f_\nu^{\mathrm{intrinsic}} =1-B_\nu(T_\mathrm{CMB}(z))/B_\nu(T_d(z)),
  \label{eq:fnucmb}
\end{equation}

\noindent
with $B_\nu$ the Planck function at rest-frame frequency $\nu$
\citep{dac13}. At $z=6.6$, this correction factor at a rest-frame
wavelength of 158\,$\mu$m is close to unity ($\sim$0.92) for a dust
temperature of $T_d=47$\,K. If we again assume a dust temperature of
$T_d=30$\,K instead of 47\,K, we already miss $\sim$25\% of the
intrinsic flux density due the CMB background emission. The fraction
of the flux density that we can measure against the CMB also depends
on the frequency and thus affects the continuum slope that we measure.

In Section~\ref{sec:j0305cont} we fitted the continuum emission of
J0305--3150 with a modified black body and derived a best-fitting
temperature in the range $T_d=30-48$\,K. If the true dust temperature
in the quasar host is in the upper end of this range, then the effects
of the CMB on our observations will be negligible, as we have
described above. On the other hand, in the case that the temperature
of the dust in J0305--3150 is closer to 30\,K, the CMB will have a
non-negligible influence on the dust properties we derive from fitting
the continuum spectrum as shown in Figure~\ref{fig:j0305fullspec}.

As a test, we therefore fitted the continuum spectrum of J0305--3150
again, this time fitting a modified black body while taking the
effects of the CMB into account: the dust temperature was modified
according to Equation~\ref{eq:tdz} and the resulting flux density was
adjusted using Equation~ \ref{eq:fnucmb}. With a fixed $\beta=1.6$ and
redshift $z=6.6145$ (Table~\ref{tab:firprop}), we measure a lower
intrinsic dust temperature of $T_d^{z=0}=30$\,K, with a 1\,$\sigma$
range of 21--42\,K. Although the dust heating due to the CMB is
negligible ($\sim$2\% temperature increase) for this source if
$T_d^{z=0}=30$\,K, observing against the CMB background reduces the
flux density we are measuring to approximately 77\% of the intrinsic
flux density.

Although our error bars are large, the lower dust temperature derived
from the continuum slope suggests that we may have overestimated the
infrared luminosity of this quasar host by assuming $T_d=47$\,K. Using
$T_d=30$\,K, $\beta=1.6$ and taking into account that the intrinsic
flux density is a factor $1/0.77=1.3$ higher than the observed flux
density, we derived an intrinsic \lfir\,$=2.6\times10^{12}$\,\lsun,
which is below the range of \lfir\ we estimated for this
quasar host ($(4.0-7.5)\times10^{12}$\,\lsun, see Section~\ref{sec:j0305res} and
Table~\ref{tab:firprop}). The total infrared luminosity \ltir\ would
also be below our previous estimates:
\ltir\,$=3.7\times10^{12}$\,\lsun, implying, if powered by star
formation, a SFR$_\mathrm{TIR}=545$\,\msunyr. We note that we have
fixed $\beta$ here; to further constrain the shape and luminosity of
the far-infrared continuum and the properties of the dust, we require
additional photometry at different frequencies (for example, continuum
measurements in other ALMA bands).

\subsection{Dynamical mass estimates}
\label{sec:mdyn}

\begin{figure}
\includegraphics[width=\columnwidth]{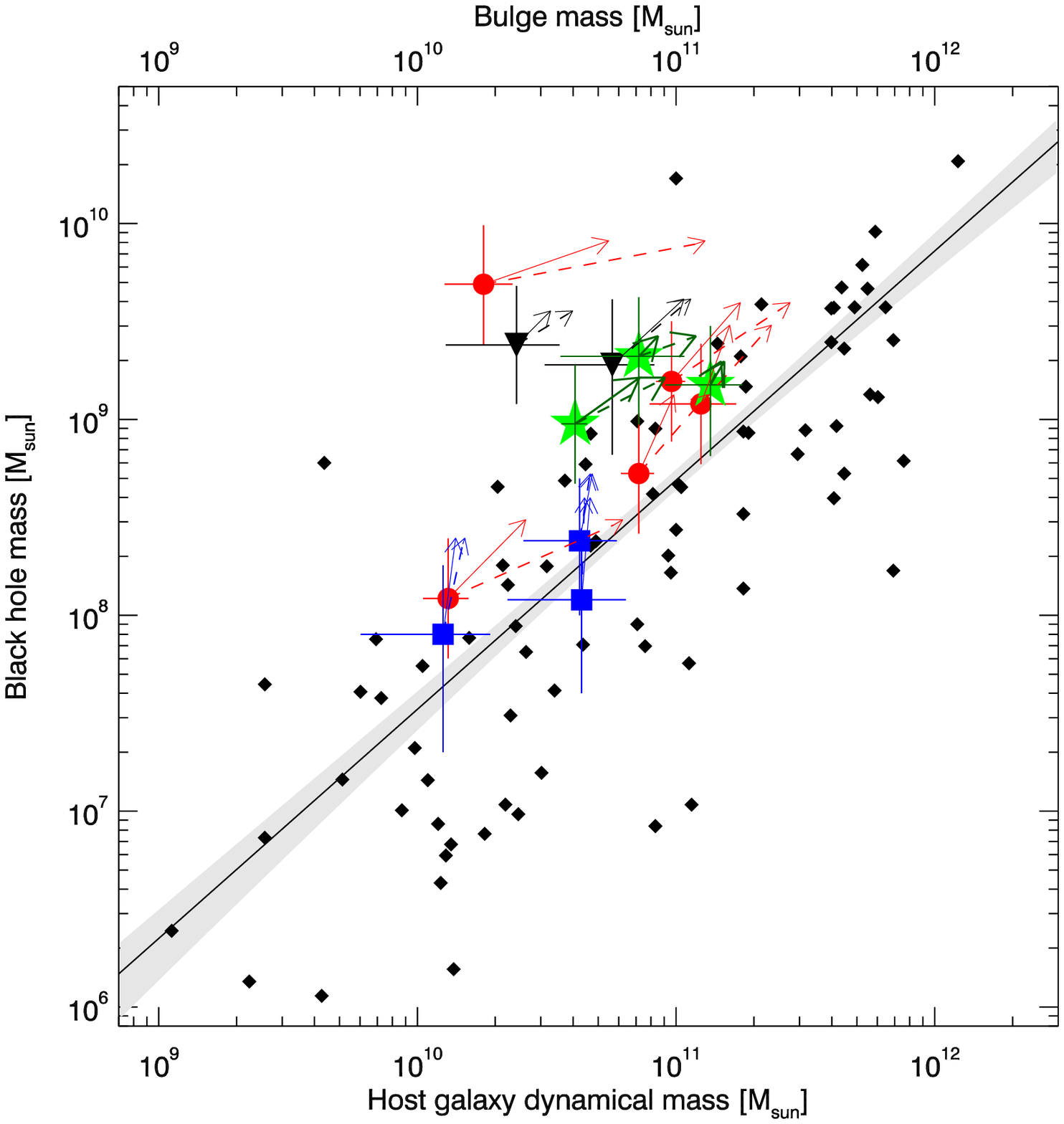}
\caption{Black hole mass plotted against the dynamical mass of
  $z\gtrsim6$ quasar host galaxies and the bulge mass of local
  galaxies. The black diamonds are values obtained for local galaxies
  (taken from \citealt{kor13}). The solid line and the shaded area
  shows the local \mbh-$M_\mathrm{bulge}$ relation derived by
  \citet{kor13}. Values for $z\gtrsim6$ quasar hosts are plotted in
  the same colored symbols as in
  Fig.~\ref{fig:lciilfir}. Following \citet{wil15}, we added
  0.3\,dex in quadrature to the errors to include the systematic
  uncertainty in deriving a black hole from local scaling
  relations. The values of the dynamical mass for the $z<6.5$ quasar
  hosts are taken from \citet{wil15}. The green stars are the $z>6.5$
  quasars presented in this work. For a given bulge mass, the high
  redshift quasars have a more massive black than local galaxies. The
  arrows indicate the black hole mass and galaxy mass of the quasar
  hosts if the measured black hole growth and SFR continue for the
  next 50\,Myr. The solid lines show the galaxy mass growth using the
  \cii-derived SFR, while the dashed lines use the \ltir\ SFR estimates.
  \label{fig:mbulge}}
\end{figure}

A procedure often applied in the literature to compute the dynamical
mass \mdyn\ in quasar hosts is to use the line width and spatial
extent of the emission \citep[e.g.,][]{wal03,wan13,wil15}:
$M_\mathrm{dyn}=1.16\times10^5 v_\mathrm{circ}^2 D$\,\msun, where
\vcirc\ is the circular velocity in \kms\ and $D$ the diameter of the
gas disk in kpc. Following \citet{wan13} we assume that the gas is
distributed in an inclined disk and the circular velocity is given by
$v_\mathrm{circ}=0.75 \mathrm{FWHM} / \mathrm{sin}(i)$
\citep[e.g.,][]{ho07} with $i$ the inclination angle. The inclination
angle can be derived from the observed minor to major axis ratio:
$i=\mathrm{cos}^{-1}(a_\mathrm{min}/a_\mathrm{maj})$, assuming a thin
disk geometry. For J0109--3047 and J0305--3150 we (marginally) resolve
the \cii\ emitting region and we derive an inclination angle of
25$^\circ$ and 50$^\circ$, respectively. For J2348--3054, we assume
the inclination angle is similar to that of other $z\gtrsim6$ quasar
hosts, which have a median inclination angle of $i=55^\circ$
\citep{wan13}. The diameter $D$ is set to $1.5\times$ the deconvolved
size of the \cii\ emitting region \citep[][see Table
  \ref{tab:firprop}]{wan13}. For the unresolved emission in the host
of J2348--3054 we assume an extent of $3\pm1$\,kpc. With these
numbers, we calculate a dynamical mass of
$(7.2\pm3.6)\times10^{10}$\,\msun, $(1.4\pm0.4)\times10^{11}$\,\msun,
and $(4.1\pm0.5)\times10^{10}$\,\msun\ for J2348--3054, J0109--3047,
and J0305--3150, respectively. The uncertainty in these dynamical
masses does not include the uncertainties in the inclination angle.

The dynamical galaxy mass we derived for each of the quasar hosts is
the sum of all the mass inside the central few kpc of the
galaxies. There are various galaxy components that contribute to this
mass: the central black hole, stars, dust and gas, and dark
matter. While the black hole contributes only a small fraction to the
dynamical mass, the gas can provide a significant fraction of the
mass. If we take the local gas-to-dust mass ratio of
$\sim$100 \citep[e.g.,][]{dra07}, the derived dust masses in our
quasar hosts of $M_d=(1-24)\times10^8$\,\msun\ imply gas masses of
$M_g=(1-24)\times10^{10}$\,\msun. In some cases, this is a significant
fraction of the computed dynamical mass for these objects. If we
assume the extreme case that the dynamical mass is all in the form of
stars in a bulge, we get an upper limit on the bulge mass in these
quasar hosts. In Fig.~\ref{fig:mbulge}, we show \mbh\ as function of
$M_\mathrm{dyn}$. We also plotted the values of the black hole mass
and bulge mass for local galaxies, which follow the relation:
\mbh/$10^9$\,\msun\,$=(0.49^{+0.06}_{-0.05})(M_\mathrm{bulge}/10^{11}
M_\odot)^{1.17\pm0.08}$ \citep{kor13}. All but one of the $z\gtrsim6$
quasar hosts are laying above the local relation. In other words, for
a given bulge (or dynamical) mass, the black holes in the high
redshift quasar hosts are more massive than those of local
galaxies. Since the true stellar bulge masses in the quasar hosts are
likely lower than \mdyn, the offsets only get more extreme. Fixing the
slope of 1.17 in the black hole--bulge mass relation, we computed the
average ratio of \mbh/\mdyn\ at a galaxy mass of $10^{11}$\,\msun\ for
the quasar hosts at $z\gtrsim6$. We find a mean of 1.9\% and a median
of 1.3\%, compared to the local value of
$0.49^{+0.06}_{-0.05}$\%. This is a factor $\sim$3--4 higher, in
agreement with studies of the host of quasars at $z\gtrsim2$
\citep[e.g.,][]{mcl06,peng06,shi06,dec10,mer10,tar12}. If we
parametrize the redshift evolution of the black hole--bulge mass
relation as
$M_\mathrm{BH}/M_\mathrm{bulge}=(M_\mathrm{BH}/M_\mathrm{bulge})_{z=0}
\times (1+z)^\beta$ \citep[e.g.,][]{mcl06,mer10,ben11,tar12}, then we
find for the $z\gtrsim6$ quasar hosts a mean of $\beta=0.7$ and a
median of $\beta=0.5$, again in agreement with the literature.

We also investigated the $M_\mathrm{BH}-\sigma$ relation by computing the
velocity dispersion $\sigma$ from the circular velocity:
$\mathrm{log}(v_\mathrm{circ})=(0.84\pm0.09)\,\mathrm{log}(\sigma) +
(0.55\pm0.19)$ \citep{fer02}. However, it remains unclear whether this
relation between $v_\mathrm{circ}$ and $\sigma$ can be applied here as
\citet{kor13} argue that the tight correlation between these parameter
might only be valid for galaxies that contain an actual bulge (which is
unknown in these quasar hosts). Nevertheless, the $z\gtrsim6$ data
point are on average above the local $M_\mathrm{BH}-\sigma$ relation,
similar to the points in Fig.~\ref{fig:mbulge}. If we calculate the
dispersion by simply converting the FWHM of the Gaussian \cii\ line to
a $\sigma$ ($\sigma$\,=\,FWHM\,/\,$(2 \sqrt{2 \mathrm{ln}(2)})$), the
$\sigma$ of the quasar hosts are smaller than those computed from
\vcirc, and the points are even further away from the local relation.

Finally, we can estimate in which directions the points in
Fig.~\ref{fig:mbulge} are moving with cosmic time. This would address
the question whether, over time, the star formation rates measured in
the distant quasar hosts will move the host galaxies on the local
\mbh--$M_\mathrm{bulge}$ relation. From the bolometric luminosity of
the quasar (see Section~\ref{sec:correlations} and
Equation~\ref{eq:lbol}), we can compute the growth of the black hole
{\it \.{M}}$_\mathrm{BH}$: {\it
  \.{M}}$_\mathrm{BH}=\frac{1-\eta}{\eta}\,\frac{L_\mathrm{bol}}{c^2}$
\citep{bar15}, with $\eta$ the radiative efficiency
\citep[$\eta\approx0.07$,][]{vol05}. For example, for the VIKING
quasars we derive a black hole growth of 11, 10, and 14\,\msunyr, for
J2348--3054, J0109--3047, and J0305--3150, respectively. When compared
to the SFRs of 270, 355, and 630\,\msunyr\ derived from the \cii\ line
(Section~\ref{sec:results}), the black hole is growing at a rate of
2--4\% of the SFR. Since these quasar hosts already have a higher
black hole to bulge mass ratio than the local value of 0.49\%, this
high growth ratio means that the quasar hosts will not move towards
the local relation over time (assuming that the accretion/growth rates
do not change). Similarly, if we assume that the galaxies are growing
with SFRs derived from \ltir\ ($\sim$950, 270, and 540\,\msunyr\ for
J2348--3054, J0109--3047, and J0305--3150, respectively, see
Sections~\ref{sec:j2348res}, \ref{sec:j0109res}, and \ref{sec:cmb}),
the black holes are growing at a rate of 1--4\% of the SFR.

In Fig.~\ref{fig:mbulge}, we show the direction of the relative growth
of black hole and galaxy mass in a time span of 50\,Myr for the
$z\gtrsim6$ quasar hosts with respect to the local black hole--bulge
mass relation. The black hole growth was computed using the bolometric
luminosity of the central AGN. The galaxy growth was calculated using
either SFR$_\mathrm{[CII]}$ (solid lines) or SFR$_\mathrm{TIR}$
(dashed lines), with the assumption that all the far-infrared emission
arises from star formation and that the material forming the new stars
is accreted onto the galaxy (i.e., that the dynamical mass were to
increase by that amount). If we assume that the SFR is traced by the
\cii\ emission, then for the majority of quasar hosts, especially the
ones close to the local \mbh-$M_\mathrm{bulge}$ relation, the black
hole is growing faster than the host galaxies: on average, in 50\,Myr
the black hole increases its mass by a factor $\sim$2.2, while the
host galaxy grows by a factor 1.35--1.55 in mass. The
\mbh/\mdyn\ ratio for the quasar hosts will therefore have increased
after 50\,Myr from 1.3--1.9\% to 2.0--2.5\%. It is possible, however,
that by using the \cii\ luminosity, we underestimated the SFR,
especially in the quasar hosts with low \lcii/\lfir\ ratios
\citep[e.g.,][]{del14}. Taking instead the TIR luminosity to derive
SFRs (which could overestimate the SFR, see e.g.,
Section~\ref{sec:correlations}), the galaxies grow by a factor of
1.7--2.3 in 50\,Myr and as a result the black hole to bulge mass ratio
will be similar at 1.3--1.9\%. Even with the larger SFR$_\mathrm{TIR}$
the vast majority (if not all) of the quasar hosts will lie above the
local relation.

\section{SUMMARY}
\label{sec:summary}

In this paper we presented short ($\sim$15\,min) ALMA observations of
three quasars at $z>6.6$: J0305--3150 at $z=6.61$, J0109--3047 at
$z=6.75$, and J2348--3054 at $z=6.89$. All three quasars have been
detected in the \cii\ emission line and in the underlying,
far-infrared continuum at high significance. 

\begin{itemize}

\item We measure \cii\ line fluxes between 1.6--3.4\,Jy\,\kms, which
  corresponds to \cii\ luminosities of
  \lcii\,$=(1.9-3.9)\times10^9$\,\lsun. This is 2--3 times brighter
  than the \cii\ line in the most distant quasar known, ULAS
  J1120+0641 at $z=7.1$, but fainter than the \cii\ line in P036+03 at
  $z=6.5$ \citep{ban15b}. The \cii\ line width are 255, 340, and
  405\,\kms, very similar to that of other $z\gtrsim6$ quasar hosts
  \citep[e.g.,][]{wan13,wil15}. For two sources, J0109--3047 and
  J0305--3150, we resolved the \cii\ line emission and we derive sizes
  2--3\,kpc, which is again similar to the sizes of \cii\ emitting
  regions in $z\sim6$ quasar hosts
  \citep[][]{wal09b,wan13,wil13,wil15}. By modeling the brightest of
  our detected \cii\ emission lines with disk models, we can rule out
  that the gas has a flat rotation curve.
  
\item From the line free channels we obtain continuum flux densities
  of 0.56--3.29\,mJy around 158\,$\mu$m (rest-frame). Depending on the
  shape of the dust continuum, the far-infrared luminosity of the
  quasar hosts is \lfir\,$=(0.6-7.5)\times10^{12}$\,\lsun. The total
  infrared luminosities are $(0.9-10.6)\times10^{12}$\,\lsun\ and we
  derive dust masses between $(0.7-24)\times10^8$\,\msun. Only in the
  case of J0305--3150, the quasar host with the brightest continuum
  emission, we spatially resolve the continuum emission, with a
  deconvolved continuum size that is smaller than the size of the
  \cii\ emitting region in the same object.

\item We fitted the slope of the FIR continuum in J0305--3150 to put a
  constraint on the dust temperature. After taking the effects of the
  CMB into account, we derive a dust temperature of
  $T_d=30^{+12}_{-9}$\,K. This is lower than the canonical value of
  $T_d=47$\,K assumed for distant quasar hosts. The FIR luminosity
  implied by a dust temperature of $T_d=30$\,K is a factor $\sim$3
  lower compared to \lfir\ computed using $T_d=47$\,K, illustrating
  that caution has to be taken when deriving FIR and TIR luminosities
  from single continuum measurements of distant quasar hosts.

\item The \cii\ equivalent widths are 0.43, 1.90, and 0.55\,$\mu$m,
  for J2348--3054, J0109--3047, and J0305--3150, respectively. These
  values are at most a factor of 3 below that of local starburst
  galaxies which have a median EW$_\mathrm{[CII]}=1.3$\,$\mu$m
  \citep{dia13,sar14}. Depending on the shape of the FIR continuum, the
  \lcii/\lfir\ range from $(0.3-4.6)\times10^{-3}$. J0109--3047 has a
  ratio of $(1.4-4.6)\times10^{-3}$, consistent with local
  star-forming galaxies. The other two quasar hosts have low values,
  $(0.3-1.0)\times10^{-3}$, similar to FIR bright quasar hosts at
  $z\sim6$ \citep{wan13}.
  
\item If the \cii\ and continuum emission are powered by star
  formation, we find star-formation rates from 140--895\,\msunyr\ for
  J0109--3047, based on local scaling relations. For the other two
  sources we derive SFRs from \lcii\ between 100--1585\,\msunyr\ and
  SFR$_\mathrm{TIR}=555-1580$\,\msunyr.

\item We combined our results with those of $z\gtrsim6$ quasars with
  \cii\ measurements in the literature. We find that the strength
  of the \lcii\ and \lfir\ emission both correlate with the bolometric
  luminosity \lbol\ of the quasar. The \lcii/\lfir\ ratio only weakly
  correlates with \lbol, implying that low \lcii/\lfir\ ratios in
  quasar hosts are not mainly due to high \lfir\ due to quasar heating
  of the dust. 
  
\item The \cii\ line in J0109--3047 is shifted by 1700\,\kms\ with
  respect to the \mgii\ line that was used to tune the ALMA
  observations. We compared the redshifts of 11
  $z\gtrsim6$ quasars based on the \mgii\ line, coming from the quasar
  broad line region, with the host galaxy redshifts traced by \cii\ or
  CO. Of these 11 quasars, 8 have a blueshifted \mgii\ line with
  respect to the host galaxy redshift. The average blueshift of the
  sample is $480\pm630$\,\kms. The \mgii\ shifts are uncorrelated
  with the luminosity and accretion rate of the central AGN, and with
  the host galaxy brightness.

\item Finally, we derived dynamical masses for the quasar hosts from
  the observed \cii\ line width and spatial extent. We find that the
  ratio of black hole mass to host galaxy mass is higher by a factor
  3--4 than local relations. We find that the black hole to galaxy
  mass ratio evolves as $(1+z)^{0.5-0.7}$, indicating that black holes
  grow faster than their host galaxies in the early universe for the
  quasars considered here. This is supported by the relative growth
  rates: we computed the growth rate of the black holes (derived from
  the quasar's bolometric luminosities) and that of the host galaxies
  (based on the measured SFRs) and, on average, the black holes are
  growing at least as fast as their host galaxies.

\end{itemize}

\acknowledgments We thank the referee for carefully reading the
manuscript and providing valuable comments and suggestions. B.P.V. and
F.W. acknowledge funding through the ERC grant ``Cosmic
Dawn''. Support for R.D. was provided by the DFG priority program 1573
``The physics of the interstellar medium''.

This paper makes use of the following ALMA data:
ADS/JAO.ALMA\#2012.1.00882.S. ALMA is a partnership of ESO
(representing its member states), NSF (USA) and NINS (Japan), together
with NRC (Canada) and NSC and ASIAA (Taiwan), in cooperation with the
Republic of Chile. The Joint ALMA Observatory is operated by ESO,
AUI/NRAO and NAOJ.

This publication makes use of data products from the {\em Wide-field
  Infrared Survey Explorer}, a joint project of the University of
California, Los Angeles, and the Jet Propulsion Laboratory/ California
Institute of Technology, funded by the National Aeronautics and Space
Administration.

{\it Facilities:} 
\facility{ALMA}.

\end{document}